\documentclass[twocolumn]{aastex63}

\usepackage{amsmath}
\usepackage{amssymb}
\usepackage{amsthm}
\usepackage{graphicx}
\usepackage{natbib}
\usepackage{subfigure}
\usepackage{multirow}

\setwatermarkfontsize{0.75in}

\begin{document}

\title{SOAR/Goodman Spectroscopic Assessment of Candidate Counterparts of the LIGO--Virgo Event GW190814}\thanks{Based on observations obtained at the Southern Astrophysical Research (SOAR) telescope, which is a joint project of the Minist\'{e}rio da Ci\^{e}ncia, Tecnologia, Inova\c{c}\~{o}es e Comunica\c{c}\~{o}es (MCTIC) do Brasil, the US National Science Foundation’s National Optical-Infrared Astronomy Research Laboratory (NOIRLab), the University of North Carolina at Chapel Hill (UNC), and Michigan State University (MSU).}


\correspondingauthor{Douglas Tucker}
\email{dtucker@fnal.gov}

\author[0000-0001-7211-5729]{D.~L.~Tucker}
\affil{Fermi National Accelerator Laboratory, P. O. Box 500, Batavia, IL 60510, USA}

\author[0000-0001-8653-7738]{M.~P.~Wiesner}
\affil{Benedictine University, Department of Physics, 5700 College Road, Lisle, IL 60532, USA}

\author[0000-0002-7069-7857]{S.~S.~Allam}
\affil{Fermi National Accelerator Laboratory, P. O. Box 500, Batavia, IL 60510, USA}

\author[0000-0001-6082-8529]{M.~Soares-Santos}
\affil{Department of Physics, University of Michigan, Ann Arbor, MI 48109, USA}

\author[0000-0003-4383-2969]{C.~R.~Bom}
\affil{Centro Brasileiro de Pesquisas F\'isicas, Rua Dr. Xavier Sigaud 150, CEP 22290-180, Rio de Janeiro, RJ, Brazil}
\affil{Centro Federal de Educa\c{c}\~ao Tecnol\'ogica Celso Suckow da Fonseca, Rodovia M\'ario Covas, lote J2, quadra J, CEP 23810-000, Itagua\'i, RJ, Brazil}

\author{M.~Butner}
\affil{East Tennessee State University, 1276 Gilbreath Dr., Box 70300, Johnson City, TN 37614, USA}

\author[0000-0001-9578-6322]{A.~Garcia}
\affil{Department of Physics, University of Michigan, Ann Arbor, MI 48109, USA}

\author[0000-0002-7016-5471]{R.~Morgan}
\affil{Physics Department, 2320 Chamberlin Hall, University of Wisconsin-Madison, 1150 University Avenue Madison, WI 53706-1390, USA}
\affil{Legacy Survey of Space and Time Corporation Data Science Fellowship Program}

\author[0000-0002-5115-6377]{F.~Olivares~E.}
\affil{Instituto de Astronom\'{\i}a y Ciencias Planetarias, Universidad de Atacama, Copayapu 485, Copiap\'o, Chile}

\author[0000-0002-6011-0530]{A.~Palmese}
\affil{Fermi National Accelerator Laboratory, P. O. Box 500, Batavia, IL 60510, USA}
\affil{Kavli Institute for Cosmological Physics, University of Chicago, Chicago, IL 60637, USA}

% ...

\collaboration{1000}{(DES Collaboration)}

\begin{abstract}
On 2019 August 14 at 21:10:39 UTC, the LIGO/Virgo Collaboration (LVC) detected a possible neutron star-black hole merger (NSBH), the first 
ever identified. An extensive search for an optical counterpart of this event, designated GW190814, was undertaken using 
{ the Dark Energy Camera (DECam) on the 4m Victor M. Blanco Telescope at the Cerro Tololo Inter-American Observatory.}
Target of Opportunity interrupts were issued on 8 separate nights to observe 11 candidates using the 
{ 4.1m Southern Astrophysical Research (SOAR) telescope's} 
Goodman { High Throughput} Spectrograph in order to assess whether any of these transients was likely to be an optical counterpart of the possible NSBH merger. 
Here, we describe the process of observing with SOAR, the analysis of our spectra, our spectroscopic typing methodology, and our resultant  conclusion that none of the candidates corresponded to the { gravitational wave merger event} but were all instead other { transients}.  
Finally, we describe the lessons learned from this effort.  Application of these lessons will be critical for a successful community spectroscopic follow-up program for LVC { observing run} 4 (O4) and beyond.

\end{abstract}


\reportnum{DES-2020-601}
\reportnum{FERMILAB-PUB-21-454-AE-E-SCD}


\section{Introduction}
\label{sec:intro}
\clearpage
\noindent 
The 2017 discovery of the optical counterpart of a binary neutron star (BNS) merger –- a kilonova { (KN)} –- was one of the highlights of observational astrophysics of the early 21st Century. This discovery, following on the 2015 discovery of the first ever detected gravitational 
wave (GW) event, GW150914 \citep{Abbott16}, was a significant leap forward for astrophysics. 
{ The detection of GW170817 in coincidence with a short gamma-ray burst by Fermi-GBM during the second observing run (O2) of the Advanced LIGO
\citep{2015CQGra..32g74001A} and Virgo \citep{2015CQGra..32b4001A} 
network inaugurated the era of multi-messenger astronomy with GWs \citep{2017PhRvL.119p1101A,2017ApJ...848L..12A}.}
The optical counterpart was discovered 12 hours after the merger by several independent teams, including our own team, the Dark Energy Survey Gravitational Wave Search and Discovery Team (DESGW). DESGW utilizes the { Dark Energy Camera (DECam) \citep{brenna} on the Victor M. Blanco Telescope at Cerro Tololo Interamerican Observatory (CTIO) in Chile \citep{2017ApJ...848L..16S}}
This discovery enabled panchromatic imaging and spectroscopy, which galvanized the astronomical community. 

While this single event captured the focus of the entire astronomical community, the breadth and number of scientific analyses stemming from it are perhaps more astounding. Standard siren techniques enabled a direct measurement of the expansion rate of the Universe today (\citealt{Abbott2017}; \citealt*{Soares_Santos_2019}; \citealt{darksiren2}) and in the future they will also be a useful probe of the growth of structure \citep{palmese20_pv}. Measuring elemental abundances in the merger ejecta using spectroscopic instruments led to an understanding of the origin of heavy elements synthesized during the merger { \citep{chornock17, Drout_2017, Tanaka_2018}, and we note the unique wavelength coverage of the VLT X-Shooter in this task in particular \citep{2017Natur.551...67P, 2017Natur.551...75S,  2019Natur.574..497W}}. X-ray and radio observations characterized the geometry of the explosion to be best described by a jet plus cocoon structure { \citep{alexander17, 2017Sci...358.1579H, margutti17, 2017Natur.551...71T, 2018Natur.561..355M, 2019Sci...363..968G}}. The gravitational waveforms tested and further bolstered the validity of the theory of General Relativity, as verified by numerical relativity simulations \citep{shibata2017, abbott2019}, and several other studies explored the connection between BNS mergers and short Gamma Ray Bursts (sGRBs) ({\em e.g.\/}, \citealt{2017arXiv171005450F,2017ApJ...848L..23F, Savchenko_2017, 2017ApJ...850L..41X, Lyman2018, Ascenzi_2020}). 
{ T}hese analyses, and many 
not listed, were enabled by the association of the GW signal with its electromagnetic { (EM)} signal. Given that these events are such a rich source of astrophysical knowledge, finding counterparts to GW events related to compact object mergers remains a primary goal of the multimessenger-focused astronomical community. 

On 2019 August 14 at 21:10:39 UTC, during its { observing run} 3 (O3), the LVC detected a binary merger initially designated as S190814bv and later given a final designation of GW190814.
This was one of 56 event alerts from LVC during O3 and was particularly interesting: GW190814 was at the time 
{ classified as} 
a neutron star-black hole (NSBH) merger, the first high significance event of this kind ever observed \citep{GCN_324, GCN_333, Abbott2020}. The LIGO-VIRGO analysis found that this merger event occurred at a  distance of $267 \pm 52$ Mpc. It had a 90\% localization region of 23 deg$^2$ and a probability of being a NSBH merger of greater than 99\%.  Further, taking as an assumption that the GW170817 BNS { KN} 
(at a distance of 43~Mpc) had a typical luminosity for such an event and scaling by the inverse-square law, 
{ one could estimate that the optical counterpart to GW190814 could conceivably peak at a brightness of $i$$\sim$21 ($\approx$4 mag fainter than that of GW170817)}
-- well within the range of DECam, as well as still within the range of 
{ medium resolution spectrographs on 4m-class optical telescopes -- simplifying the effort of following up any likely optical counterpart candidates.}
Thus, the DESGW team undertook an extensive search for a { KN} 
event that would form the optical counterpart to this potential NSBH merger event, making use of DECam observations within the high-probability region of the GW event. This search is described in detail in 
\citet{Morgan_paper}. 

A number of other groups also searched for an { EM} counterpart to GW190814.  \citet{Kilpatrick_2021} (many of whom are also members of the DESGW Collaboration) discuss searches for { KN} 
candidates using several 0.7-1 meter class telescopes as well as Keck/MOSFIRE and also present spectroscopy of a number of candidates { (including in their Figure~4 a copy of many of the spectra described here in the current paper).} They also present limits on { EM} counterparts to GW190814 and consider scenarios in which an { EM} counterpart of a NSBH would be detected. The Australian Square Kilometre Array Pathfinder (ASKAP) imaged 30~deg$^2$ at 2, 9 and 33 days after the event at a  frequency of 944 MHz \citep{Dobie19}. The Magellan Baade 6.5 m telescope was used to search { on} a selection of galaxies within the localization area out to limiting magnitude of $i=22.2$ and found no counterparts \citep{Gomez19}. The MegaCam instrument on the Canada-France-Hawaii Telescope (CFHT) was used to search much of the localization region. Although the CFHT team reached a depth of $i>23.9$ at 8.7 days post-merger, no { KN} 
was found \citep{Vieira20}. The GROWTH Collaboration used imaging from DECam along with other facilities for imaging and spectroscopy of possible { KN} 
candidates.  Using simulations, they constrained possible ejecta mass from the merger to be $M_{\rm ejecta}<0.04$ $M_{\sun}$ at polar viewing angles \citep{Andreoni19}. \citet{Watson20} described limits on an { EM} counterpart to GW190814 using observations with optical imager DDOTI (at the Observatorio Astron\'{o}mico Nacional in Mexico) and Swift/BAT observations. They showed that Swift/BAT should have detected an associated gamma ray burst at the 98\% level. \citet{Ackley20} described the ENGRAVE team search using the Very Large Telescope as well as involvement with the ATLAS, GOTO, GRAWITA-VST, Pan-STARRS and VINROUGE projects. 
Their observations covered the localization region to depths as faint as $r\approx 22$.
Their limits suggest that it is likely the neutron star was not disrupted during the merger. DDOTI wide-field observations were also used along with the Lowell Discovery Telescope, the Reionization and Transients InfraRed and spectroscopy from the Gran Telescopio Canarias to locate { EM} counterparts \citep{Thakur20}. Their data suggest that there was no gamma ray burst along the jet's axis. 

While searching for an optical counterpart to GW190814, the DESGW pipeline began with 33,596 events in the likelihood regions. Using the analysis pipeline we produced a final list of 11 candidates that passed our cuts and were bright enough for spectroscopy using a 4-m class telescope (\citealt{Morgan_paper}; also \S~\ref{sec:filter} below). For these candidates we proceeded to conduct spectroscopic typing at the Southern Astrophysical Research (SOAR) 4.1\,m telescope\footnote{\url{https://noirlab.edu/science/programs/ctio/telescopes/soar-telescope}} using the Goodman High Throughput Spectrograph (HTS; \citealt{Clemens2004}). (Spectroscopic typing is facilitated by the fact that, due to the fast ejecta velocities expected of kilonovae --- 0.03-0.30$c$ --- their spectra are expected to be featureless or only have very broad, smooth spectral features, especially in the optical during the first few days after the merger event, which distinguishes their spectra from supernovae { [SNe]} and other optical transients; see, e.g, the
{ KN} models of \citealt{kasen_2017}.)  The spectroscopic follow-up team submitted Target of Opportunity (ToO) observing requests to the SOAR telescope on 8 separate nights in order to use the Goodman { HTS} on SOAR for spectroscopic typing of these 11 candidates.  

After taking spectra for 8 candidates (plus the host galaxies of 3 additional candidates which had faded beyond the straightforward capabilities of SOAR -- {\em i.e.\/} $i\sim21.5$), no optical counterpart was discovered for GW190814. 
{ Despite this null result, this paper serves several important functions. First, it serves as a companion paper to our other two papers \citep{Morgan_paper, Kilpatrick_2021}, providing a deep dive into the methodology and detailed results of a coordinated spectroscopic campaign of the first possible NSBH event ever detected, including the finding charts, light curves, and KN spectral fitting not covered in detail by the other two companion papers.  Further, it describes and provides previously unpublished open source tools that can be of use to similar future spectroscopic campaigns. Also, by comparing results from two separate SN spectrum fitters and a KN spectrum fitter, this paper goes into some detail into the subtleties associated with spectroscopic classification of relatively faint SNe and KNe.  Finally, although it does not change the conclusions of the companion papers, some of the final classifications of the candidate counterparts here are updates from what what was seen in the previous papers.  }




{ In summary, we describe in this paper the DESGW collaboration's spectroscopic follow-up campaign for the GW190814 gravitational merger event.}  We also describe our overall spectroscopic follow-up methods and strategy, how we employed them in this particular follow-up campaign, the lessons learned, and the prospects for the future. 
{ The} paper is organized as follows: In \S \ref{sec:lvc_obs} we describe the LIGO/Virgo observations of GW190814. In \S \ref{sec:decam_search} we describe the DESGW search for candidate optical counterparts. In \S \ref{sec:spectro_candidate_select} we describe the selection and filtering of the candidates. In \S \ref{sec:soar_obs} we describe the SOAR observing strategy and the observations of counterpart candidates for GW190814. In \S \ref{sec:results} we discuss our results and address the population of objects we found. In \S \ref{sec:conclusions} we summarize our conclusions. 
In addition, we provide in \S~\ref{sec:software} a list of software packages used throughout our analysis. 

In this paper we follow the cosmology given by \citet{Bennett2014}, with flat $\Lambda$CDM cosmology with $\Omega_{\rm M}=0.286 \pm 0.008$ and $H_{\rm 0}=69.6\pm 0.7$~km~s$^{-1}$~Mpc$^{-1}$.

\section{LIGO/Virgo Observations}
\label{sec:lvc_obs}
As noted above, on 2019 August 14 UTC, the LVC observed gravitational radiation at high statistical significance. The event, initially named S190814bv,
occurred during a time that all three detectors (LIGO Hanford, LIGO Livingston, and Virgo) were operating normally, which enabled both 
a good angular localization 
of the source and { more precise estimate of the source parameters}.
The false alarm probability was calculated at $2.0\times 10^{-33}$~Hz --- or once per $10^{15}$~Hubble times ---  suggesting a very high signal-to-noise event \citep{GCN_333}.  
Using the {\tt bayestar} pipeline \citep{2016PhRvD..93b4013S}, the LVC team localized the source of the { GW} signal to a $38$ $(7)$~sq.~degree area at the $90\%$ ($50\%$) confidence level in the Southern Hemisphere on the night of the merger. The initial luminosity distance estimate was $276 \pm 56$ Mpc \citep{GCN_324}. Preliminary source classification via a machine-learning-based tool \citep{2020CQGra..37d5007K} identified the event as a ``mass-gap'' binary merger -- i.e., a merger event in which at least one of the compact objects has a mass falling within the hypothetical mass gap between neutron stars { (NSs)} and black holes { (BHs)} { (i.e., in the mass range 3-5~$M_\sun$; \citealt{LIGO_user_a,abbott2020_ref})}. 
{ The small localization area and the potential of identifying an optical counterpart made this event interesting from the perspective of follow-up 
projects.}

The following day, the LVC \texttt{LALInference} pipeline \citep{lalsuite} localized the source to $23 (5)$~sq.~degrees at the $90\%$ ($50\%$) confidence level, refined the classification to { an NSBH} merger, and estimated the luminosity distance of the event to be $267\pm52$~Mpc ($z=0.059\pm0.011$ for a standard $\Lambda$CDM cosmology;  \citealt{Bennett2014}, \citealt{Wright2006}). 
S190814bv 
{ thus became the first possible NSBH system observed} by a { GW} observatory and a prime target for follow-up by the { EM} astronomical community. 
However, the LVC parameter estimation indicated that the parameter \texttt{HasRemnant} was $ < 1\%$. (\texttt{HasRemnant} is the probability that a nonzero { mass was ejected during the collision and remains outside the final remnant object [\citealt{2018PhRvD..98h1501F,LIGO_user}])}.  This suggested that there was a low probability that any ejecta was preserved outside the { BH} and thus that there was a small chance of there being an observable 
{ KN}.

Well after searches for an { EM} counterpart were completed, the LVC published results from an updated offline analysis \citep{Abbott2020}, where the final luminosity distance was estimated to be $239^{+41}_{-45}$ Mpc (median and 90\% credible interval), the 90\% localization area was updated to 18.5 square degrees, and the masses of the two merging objects was updated to 23.2~$M_{\sun}$ (a { BH}) and 2.6~$M_{\sun}$ (a mass-gap object -- {\em i.e.\/}, either an underweight { BH} or an excessively massive { NS}).  It was also at this time that { this GW event was re-named from its initial designation, S190814bv, to GW190814}.

{ The nature of this GW190814 was recently debated and summarized by \citet{Abbott2020}, and, since its discovery, only a couple more GW merger events with comparable properties have been identified (see \citealt{2021arXiv211103606T}) and the interactive plot at \url{https://ligo.northwestern.edu/media/mass-plot/index.html}).  Particularly striking is the mass ratio of the GW190814 merger components --- a value of 0.112 ---  whereas the average mass ratio of more typical LIGO BBH events is $\sim$1.  As noted above, one of the components of the GS190814 merger was a 23.2~$M_{\sun}$ BH, but the other was a 2.6~$M_{\sun}$ ``mass-gap'' object.  If this mass-gap object is an NS, this has ramifications for the NS equation of state, which is a determining factor in the maximum allowable mass of NS's (currently estimated to be $\la$2.6$M_{\sun}$).  Independent of whether the mass-gap object is a NS or a BH, if these types of mergers are more common than expected, there may be consequences for stellar population synthesis models, since these models tend to favor the merger of systems with components that are less asymmetric in mass, although stellar environment may also play a role:  merger rates between NS's and BH's are low in globular clusters \cite[$\sim$ 10$^{-2}$-10$^{-1}$ Gpc$^{-3}$ yr$^{-1}$; e.g., ][]{ye2020}, but likely higher in young stellar clusters \cite[$<$ 10$^{-1}$ Gpc$^{-3}$ yr$^{-1}$;][]{ziosi2014}; thus, star clusters with young stellar populations might be the preferrred location for mergers similar to GW198014.  For the purposes of this paper, we will assume that GW190814 is a possible NSBH merger, as it was classified during the SOAR follow-up observing runs.}

In the next section we describe the efforts of the DESGW { Collaboration to identify transients that were possible KN candidates.}

\section{DECam Search Campaign} 
\label{sec:decam_search}
In searching for an optical counterpart to GW190814, the DESGW collaboration triggered 
ToO observations with the { 570-mega pixel DECam optical imager on the CTIO Blanco 4-m telescope.}
Together, the Blanco and DECam reach a $5\sigma$ limiting $r$-band magnitude of $\sim23.5$ in a 90 second exposure in a 3 square degree field of view (FoV) \citep{des_strategy}.
The combination of deep imaging and a wide FoV make Blanco/DECam the ideal 
instrument for efficiently detecting
optical transients localized to tens of square degrees.

Our follow-up efforts for GW190814 utilized the resources of the Dark Energy Survey (DES), which is a wide-field optical survey that covered a 5,000 square  degree region
of the southern sky from 2013 to 2019 using Blanco/DECam \citep{Diehl:2019}.
DES imaging of the DES footprint reaches a $10\sigma$ depth for point sources of $grizY$ $=$ 25.2, 24.8, 24.0, 23.4, 21.7 mag \citep{des_cal}.
The LVC 90\% containment region for GW190814 is entirely within the DES footprint, enabling the use of high-quality DES images during difference imaging.

We performed DECam ToO follow-up observations of GW190814 for six nights following the LVC alert, namely nights 0, 1, 2, 3, 6, and 16. 
The early nights were chosen to look for rapidly evolving transients immediately following the merger. 
{ KNe} 
from either { BNS} \citep{arcavi17} or NSBH \citep{kawaguchi2016} events are expected to vary by about a magnitude over the course of a single night in the first days after the event. Observations 16 nights after the merger were used to exclude persisting { SNe}. Due to moon brightness, especially during the first nights of DECam follow-up, we opted to use the redder $i$ and $z$ bands to minimize the effect of sky brightness on our imaging depth.

The DECam images were processed by the DES Difference Imaging Pipeline \citep{details}, an updated version of the DES { SN} Program's Pipeline described in \citet{diffimg}, using coadded DES wide-field survey images \citep{2018ApJS..239...18A} as templates. 

After image processing, candidate { KNe} 
were identified and then selected for spectroscopic follow-up.  The selection  process included eliminating moving objects (e.g., asteroids), known transients (e.g., variable stars and { active galactic nuclei [AGN]}), and transients with colors and/or light curves characteristic of SNe.
Visual inspection of the images was also important, especially in the first nights of DECam follow-up, when light curves for the candidates consisted of only one or two epochs.  For GW190814 in particular, there were 33,596 candidates immediately after the image processing.  
{ KN} candidates were found in DECam images after running them through the reduction pipeline.  Objects were found by SExtractor \citep{sextractor}. Objects that had good detections in SExtractor, showed evidence of being transients by comparison to known object templates and passed visual inspection checks were considered. Other candidates were identified in alert notifications from the Gamma-ray Coordinates Network (GCN)\footnote{\url{https://gcn.gsfc.nasa.gov/}}
put out by other groups searching for 
kilonova 
{ KN} candidates. A more rigorous process of object assessment was done later,  
described in more detail in \citet{Morgan_paper} and summarized in \S~\ref{sec:filter}. 
In the end, spectroscopic follow-up was performed using the SOAR Goodman { HTS} for 11 candidates (or their host galaxies). 

In Table \ref{tab:candidates1} we present candidates found and spectroscopically targeted by the DESGW team during DECam follow-up of GW190814. In this table we provide both the DESGW ID and the Transient Name Server name, which we continue to use in this work. In the final two columns, {we present the localization probability enclosed within the GW sky-map including each object location.} For further details of the processing of the DECam data and the subsequent identification of possible candidates, please refer to our companion paper \citep{Morgan_paper}.  

\begin{deluxetable*}{ccccclccc}
\tablewidth{0.8\textwidth}
\tabletypesize{\scriptsize}
\tablehead{
\colhead{DESGW} &  \colhead{TNS} & \colhead{RA(2000)} & \colhead{Dec(2000)} & \colhead{GCN / ID} & \colhead{Mag at} & \colhead{band}&\colhead{Prob reg}&\colhead{Prob reg}\\
\colhead{ID} &  \colhead{Name} & \colhead{(deg)} & \colhead{(deg)} & \colhead{} & \colhead{discovery} & \colhead{}&\colhead{initial}&\colhead{final}
} 
\startdata
624921 & 2019nqq & 20.95506 & -33.034762 & 25373 / c &  20.76 & i & 90\% & o \\
624609 & 2019nqr & 23.573539 & -32.741781 & 25373 / d &  18.34 & i & 80\% & 90\%\\
624690 & 2019noq & 12.199493 & -25.30652 & 25356 (Pan-STARRS) & 19.93 & i & 30\% & 30\%\\
624157 & 2019ntn${\dagger}$ & 23.722184 & -31.380451 & 25393 (GROWTH) & 20.8 & i & 90\% & o\\
626761 & 2019npw &13.968327 & -25.783283 & 25362 / e & 20.5 & i & 40\% & 60\% \\
631360 & 2019num & 13.881714 &-22.968887 & 25393 (GROWTH) & 21.3 & i & 90\% & o \\
661833 & 2019ntr&15.007796 & -26.714266 & 25393 (GROWTH) & 21.2 & z & 80\% & o \\
625839 & 2019omx &24.18436 & -33.302678 & 25486 / z & 22.1 &  z & 90\%& o\\ 
626956 & 2019ntp & 12.550247 & -26.197878 & 25393 (GROWTH) & 21.0 & i & 50\% & 60\% \\
631484 & 2019nte & 23.557358 & -31.721700 & 25398 / f & 20.95 & i & 80\% & o\\
635566 & 2019omw & 12.234396 & -23.170137 &  25486 / y & 22.8 & i & 50\% & 80\% \\ 
\enddata
\caption{Candidates found by the DESGW team during the DECam Follow-up of GW190814 that were then followed up with SOAR ToO observations. The DESGW ID is the internal identification number while the TNS name comes from the Transient Name Server (\url{https://wis-tns.weizmann.ac.il}). The coordinates are given here in degrees, along with the GCN announcing discovery of the transient. Magnitude at discovery is given in the band listed. {The confidence probability enclosed within the GW sky-map including the object position is given both for the initial map issued by LVC used during observing and for the final, smaller map. (The ``o'' means outside the the 90\% sky-localization probability region.)}} 
\label{tab:candidates1}
\tablenotetext{\dagger}{AT2019mbq was accidentally targeted for SOAR spectroscopy instead of the intended target AT2019ntn, and this accident was not discovered until much later. {This mistake has been traced to a copying error during the handoff of this target from the DECam processing \& analysis team to the SOAR observing team.} Candidate AT2019mbq is at RA=10.835384~deg  DEC=-25.883880~deg, with a magnitude at discovery of $i=18.75$.  We note that AT2019mbq was not originally considered for spectroscopic follow-up since its host galaxy had a too high estimated photo-$z$ ($z_{\rm photo} = 0.17 \pm 0.05$) and since there was evidence of a pre-merger detection for this candidate.  As for AT2019ntn, although no spectrum was taken of it, the fact that it brightened in $z$-band about 4 days after the merger and the fact that it lay outside the 90\% confidence contour of the LVC final map (Fig.~\ref{fig:prob_map}) make it unlikely that AT2019ntn was the optical counterpart.}
\end{deluxetable*}

\begin{figure*}
\vspace{-0.5cm}
\includegraphics[width=0.90\textwidth]{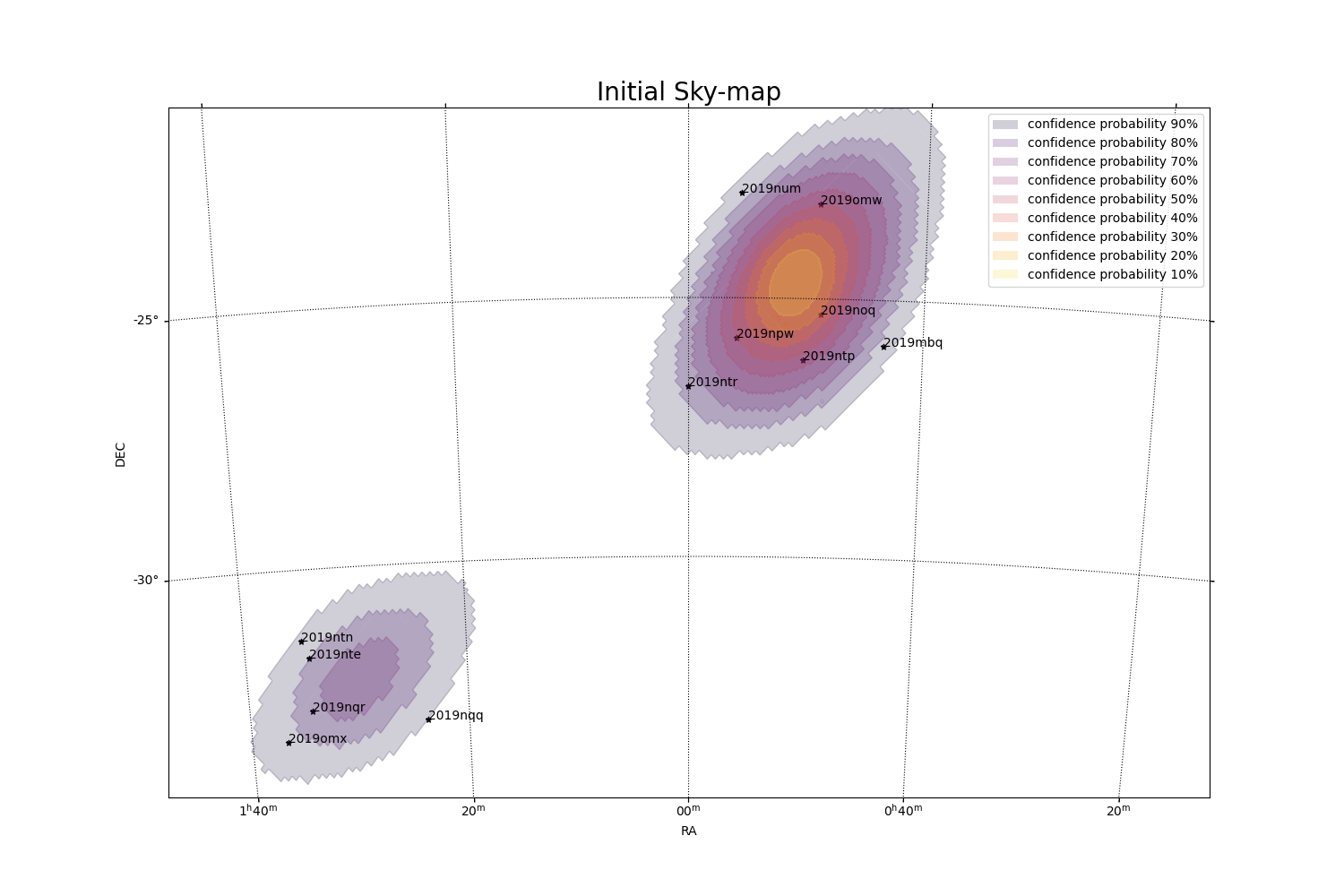}
\vspace{-1.0cm}
\includegraphics[width=0.90\textwidth]{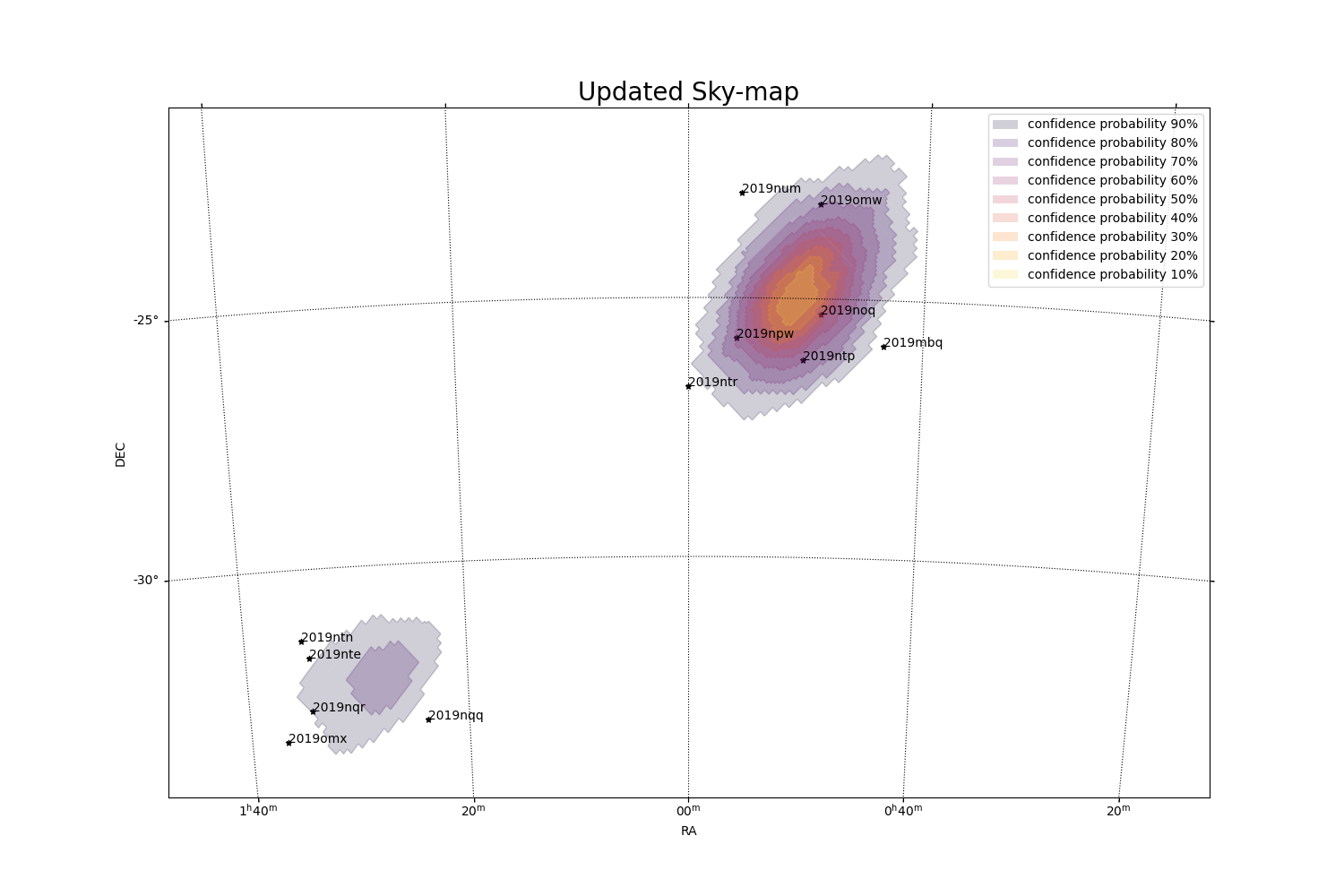}
\vspace{0.25cm}
\caption{{LVC sky-localization maps for GW190814; colors indicate confidence probability contours.} The {\em top\/} figure is the {initial sky-map}, released shortly after event discovery on 2019 August 14. The {\em bottom\/} figure is the final sky-map, released after further analysis by the LVC collaboration. The locations of each of the 11 objects we describe in this paper are also given.} 
\label{fig:prob_map}
\end{figure*}

In Figure \ref{fig:prob_map} we show both the {initial} and the final {sky-localization maps} issued by the LVC 
along with the locations of each of the 11 objects we observed.  Note that in the smaller final probability regions, some of the objects we observed {are outside the 90\% probability area, but all were included within this area in the initial map.}

\section{SOAR Spectroscopic Candidate Selection}
\label{sec:spectro_candidate_select}
To achieve the maximum science, rapid spectroscopic follow-up of candidate { KNe} 
is a necessity:  first to discover the optical counterpart from among the list of potential candidates, and then, if discovered, to permit the longest possible timeline for optical monitoring of the evolution of the potential
{ KN's} 
light curve and spectral energy distribution before it fades to obscurity.  The constraints for our 
SOAR spectroscopic program, however, were two-fold:  (1) to preserve  each night's main program as much as possible, as SOAR ToO interrupts are limited to 2.5~hours per night (including overheads); and (2) to achieve reasonable $S/N$ 
($\ga$5-10) of a medium-resolution spectrum on SOAR within a reasonable amount of time.  Due to these constraints, 
{ each observation is} limited to objects with brightnesses of $i<21$.  (We pushed the limits for { GW190814}, relaxing this constraint to $i\la21.5$.) In \S \ref{sec:baseline} we present our baseline strategy for SOAR/Goodman spectroscopy in LVC O3. Then in \S \ref{sec:filter} we describe our strategy for filtering transients found with DECam observing to find the candidates that should be followed up with spectroscopy.

\subsection{SOAR Program Baseline Strategy for LVC O3} \label{sec:baseline}

We designed our SOAR ToO program for rapid and robust identification and subsequent nightly follow-up of 
{ KN} candidates to be coupled with the DECam wide-field search \& discovery program \citep{2017ApJ...848L..16S,details,Morgan_paper,Garcia2020}, which would be providing a selection of candidates for spectroscopy. This project was awarded time at the SOAR/Goodman HTS to observe GW optical candidates discovered during the entire year-long O3 
{ run}
of the LIGO/Virgo campaign. Due to the transient nature of GW optical counterparts 
{ (KNe), } 
SOAR spectroscopy must be carried out in ToO mode. 
We requested SOAR/Goodman HTS ToO time in instant activation mode for a total of 10\,h or at least 4 ToO activations per semester. This way we took advantage of the fast survey confirmations from the DECam search \& discovery program, which could be available within 1\,h, if the merger happened during the Chilean night.  { The LVC predicted that there would likely be roughly 8 BNS mergers and 1 NSBH mergers -- the events most likely to yield an optical counterpart -- over the course of the LVC O3 run \citep{2017PhRvL.119p1101A,2017arXiv170908079C}.  Thus we planned to use SOAR to follow up the 2--3 of these events likely visible from the Southern Hemisphere each observing semester.}


The { KN} 
for the GW170817 BNS merger was exceptionally bright and easy to identify. It was expected that future events would on average be much farther away and thus likely to be much fainter and harder to distinguish from other transients (e.g.\ SNe Ia) in the larger volume encompassed by LVC O3 detection thresholds. We planned to use the { SOAR} Goodman HTS (1) to spectroscopically identify the optical counterpart to the GW event from among a small list of candidates provided by an initial DECam search \& discovery program; (2) once identified, to obtain a higher-S/N optical spectrum of the counterpart, suitable for detailed modeling; and (3) to obtain additional high-S/N spectra of the potential 
{ KN} 
on successive nights until it was effectively too faint for useful follow-up on SOAR. We would employ an instrument setup almost identical to that of \citet{nicholl17}, who were able to follow the GW170817 
{ KN} 
event at reasonable $S/N$ using the Goodman HTS from day 1.5 to day 7.5 after the GW trigger. In that case the kilonova faded from magnitude $i \approx 18$ to 21 over 6 days; they used an integration time (IT) of $3 \times 900$\,s with the 400 l/mm grating. Based on their Goodman spectra, we anticipated that we could achieve the S/N necessary to classify whether a given candidate was a true 
{ KN} 
or just another transient using a single 900\,s exposure for $i \le 19$ candidates, a single 1200\,s exposure for $i \approx 20$ candidates, and a single 1800\,s exposure for $i \approx 21$ candidates. We would leave fainter candidates to programs on larger telescopes, like programs on VLT and Gemini-South. 

\begin{figure}
\includegraphics[width=\columnwidth]{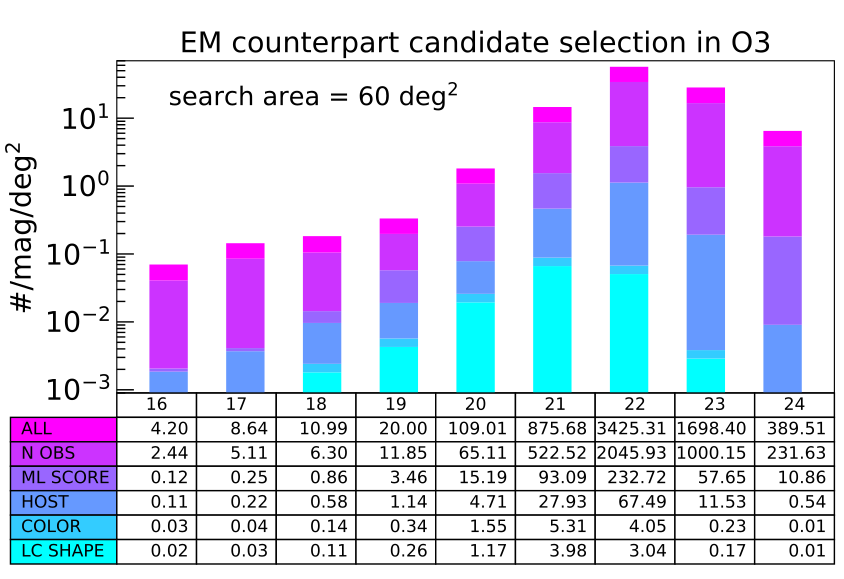}
\caption{The baseline DECam search \& discovery candidate selection for spectroscopic follow-up for LVC O3. 
The need for a robust classification pipeline to find 
{ KNe} in O3 --- as was uniquely done for GW170817 in \cite{2017ApJ...848L..16S} --- is shown here in the ($i$-band) magnitude distribution of all transient candidates expected to be found by a DECam search \& discovery imaging sequence for a typical BNS GW trigger in LVC O3, assuming a typical search area of 60~sq~deg (e.g., see \citealt{2018ApJ...852L...3S}).  { The first row (``{\tt ALL''}), which corresponds to the magenta histogram, is the distribution of candidates expected to be output from the DECam Difference Imaging Pipeline.  In these simulations, we rejected moving objects and artifacts by requiring $>$2 observations (``{\tt N\_OBS''}) and machine learning  classification score $>$0.7 (``{\tt ML\_SCORE''}), rejected  candidates with host galaxies at $z>0.2$ (``{\tt HOST''}), and performed a color cut using the fact that, unlike SNe, the early evolution of a KN is black body-like (``{\tt COLOR''}); as detection of a rising light curve would immediately pin-point the target, we applied a reduction of 25\% assuming that, given DECam scheduling constraints, we would be able to get 2 epochs at $<$24h from merger for 1 in 4 events (``{\tt LC\_SHAPE}'').  Thus, this} last row (``{\tt LC\_SHAPE}''), which corresponds to the cyan histogram, is the expected distribution of candidates remaining after all the image-level culling procedures have been run.   (Note:  the numbers listed below the plot are the total per magnitude bin for the full 60~sq~deg search area; the $y$-axis of the plot, however, is the number per magnitude bin {\em per square degree}.  Also note:  the results shown in the above plot and histogram are based on multiple simulations covering areas larger than 60~sq~deg; scaling to a 60~sq~deg localization area and averaging over the multiple simulations means that the numbers in these bins are not integers [e.g., why the number of candidates in the $i=21$ bin in the ``{\tt ALL}'' row is 875.68 and not, say, exactly 875].)}
\label{fig:candidates_o3}
\end{figure}

We planned following up the list of candidates until we either finished the list (finding no 
{ KN}) or identified the optical counterpart. For an identified 
{ KN}, two additional exposures of the same integration time would allow us to build $S/N$ suitable for model fitting.  We planned  continued SOAR follow-up if a confirmed 
{ KN} was brighter than $i=20$\,mag, requesting interrupts on all successive nights until it faded below 
{ that value}.  
We ran 100,000 simulations of the SOAR search program. An average of 8.79 DECam candidates per LIGO event in the magnitude range $i=16$--24 was assumed, where magnitudes were drawn randomly from the expected candidate distribution (see the \texttt{LC\_SHAPE} row of Fig.~\ref{fig:candidates_o3}, where the numbers add up to 8.79).  To estimate the time needed, we included not only the expected exposure times, but also all relevant overheads (e.g., slewing, target acquisition, readout, standard star observations, etc.).  To compensate for possibly worse sky transparencies (\citealt{nicholl17} found clear skies), the science integration times were multiplied by a factor of 1.25. The simulations showed that, for a single GW event, 50\% of the time a SOAR follow-up would be completed in 4.3\,h (2 ToO interrupts), 95\% of the time in 6.7\,h (3 interrupts), and 100\% of the time in 9.5\,h (4 interrupts).  Note that follow-up completion does not necessarily mean a guaranteed identification of the optical counterpart:  it may just mean that the list of candidates bright enough to be observed by SOAR was exhausted without identifying the optical counterpart or even that the optical counterpart (if any) was too faint to be detected by the DECam imaging.  Nonetheless, in our time requests, we estimated approximately 10\,h per GW event to 
optimize our chances of spectroscopically identifying and monitoring a 
{ KN} with SOAR during the LVC O3 
{ run.}

For spectroscopic classification, it was anticipated SOAR could go as faint as $i=21$. In Figure \ref{fig:candidates_o3} we visually represent the process for DECam search \& candidate selection for spectroscopic follow-up. This figure shows the expected number of DECam candidates per magnitude per square degree in LVC O3, for a typical localization area of 60 sq deg. The columns are arranged in order of magnitude, with magnitude getting dimmer to the right. 

For continued monitoring of the evolution for the optical spectrum of an identified 
{ KN}, it was thought that a higher $S/N$ would be required; so additional monitoring was planned to be constrained to 
{ KNe} brighter than $i=20$.  Candidates fainter than $i=21$ and confirmed 
{ KNe} fainter than $i=20$ would be handed over for larger telescopes for spectroscopic follow-up.
Via simple timing simulations, we estimated the amount of time to obtain SOAR spectra for typical 
{ KN} candidates from a given LVC O3 event to take no more than $\approx$10 hours over the course of $\la$5 nights (recalling the maximum ToO ``interrupt'' time per night is 2.5~hours)
The SOAR team would meet with the DECam team once the DECam team had a set of candidates.

\begin{figure*}[ht]
\centering
\includegraphics[angle=0,width=1.0\linewidth]{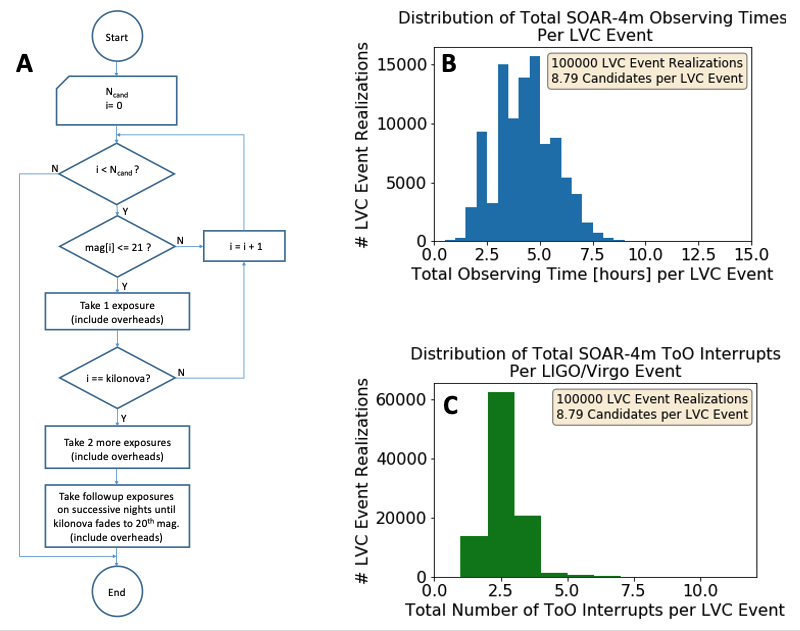}
\vspace{-.5cm}
\caption{{\em (A)\/} { A simplified flow-chart for a single realization of a simulated SOAR follow-up of a single GW event, where $N_{\rm cand}$ is the total number of candidates from an imaging search \& discovery program.\/}  For the simulations here, $N_{\rm cand}$ is either 8 or 9, but averages overall to 8.79.  The distribution of $i$-band magnitudes for the candidates is drawn from the ``LC\_SHAPE'' row in Fig. \ref{fig:candidates_o3}, and the overall average number of candidates (8.79) is just the sum of the entries in the ``{\tt LC\_SHAPE}'' row.
{\em (B)\/} Results of the simulation (using 100,000 realizations):  histogram of the total durations of SOAR ToO interrupt time [in hours] for a single LVC O3 event. 
{\em (C)\/} Results of the simulation (using 100,000 realizations):  histogram of the total number of SOAR ToO interrupts for a single LVC O3 event.  (Note that the number of interrupts does not scale exactly as the total duration of interrupt time, since the number of hours per interrupt will vary between the ``search \& discovery'' phase and the follow-up phase of the observations for a given 
{ KN} event.)}
\label{fig:soar_sims}
\end{figure*}

{ To elaborate, i}n Figure \ref{fig:soar_sims}, panel A, we present a simplified flow chart for a simulated SOAR follow-up { for the optical counterpart} of a { single LVC O3} event. $N_{\rm cand}$ is the total number of candidates from an imaging search and discovery program --  {\em i.e.\/} the expected number of objects for which we would need to take spectroscopy from SOAR or, for fainter candidates, from other telescopes.  { If we run this flowchart over 100,000 realizations and compile the results, we get the histograms in panels B \& C of Figure \ref{fig:soar_sims}.  Panel B shows the distribution -- over 100,000 simulated realizations -- of the total duration (in hours) of SOAR ToO interrupt time expected for a single LVC O3 event.  Likewise, panel shows the distribution over 100,000 simulated realizations of the total number of SOAR interrupts expected for a single LVC O3 event.} 


\subsection{Candidate Filtering for GW190814} \label{sec:filter}


For GW190814, we selected targets for SOAR spectroscopy by reducing the DECam images in real-time and monitoring the GCN for objects of interest detected by other follow-up teams.
In both approaches, one important constraint is the brightness of the candidates. 
For accurate spectroscopic classification, we wanted a minimum SNR of 5--10 in the collected spectra.
Therefore in typical observing conditions, with 45 minute to 1 hour exposure times, objects fainter than $21.5$~$i$-band~mag are excluded. 
However, if the candidate's host galaxy was brighter than the magnitude threshold, we targeted the host to obtain a precise redshift of the candidate.\footnote{We note that the host galaxy for each candidate was identified by matching the candidate's coordinates with the DES Y3 galaxy catalog using both angular and galaxy photo-$z$ information.  Details can be found in \S~3.3 of \citet{Morgan_paper}.}

The candidate selection performed in real-time for the SOAR targets differs from the offline candidate selection presented in \citet{Morgan_paper}.
One important difference is that all potential SOAR targets were selected before we began co-adding the DECam images within the same night and filter.
The cuts applied to select spectroscopic targets were:

\begin{enumerate}
    \item {\textit{ALL}. Detected in DECam images by the DESGW Search and Discovery Pipeline;}
    \item{\textit{DETECTED 2x}. At least two detections by \texttt{SExtractor} with no errors and with an \texttt{autoscan} score of at least 0.7 separated by at least one hour (\texttt{autoscan} is a machine learning-based tool for differentiating between image artifacts and real objects \citep{Goldstein15});}
    \item {\textit{PHOTO z}. If a host-galaxy exists in the DES Catalog, the estimated photometric redshift and its error must be consistent with the LVC distance mean within three standard deviations;}
    \item {\textit{INSPECTION}. Pass visual inspection by the DESGW team.}

\end{enumerate}

\begin{figure}
\centering
\includegraphics[width=\columnwidth]{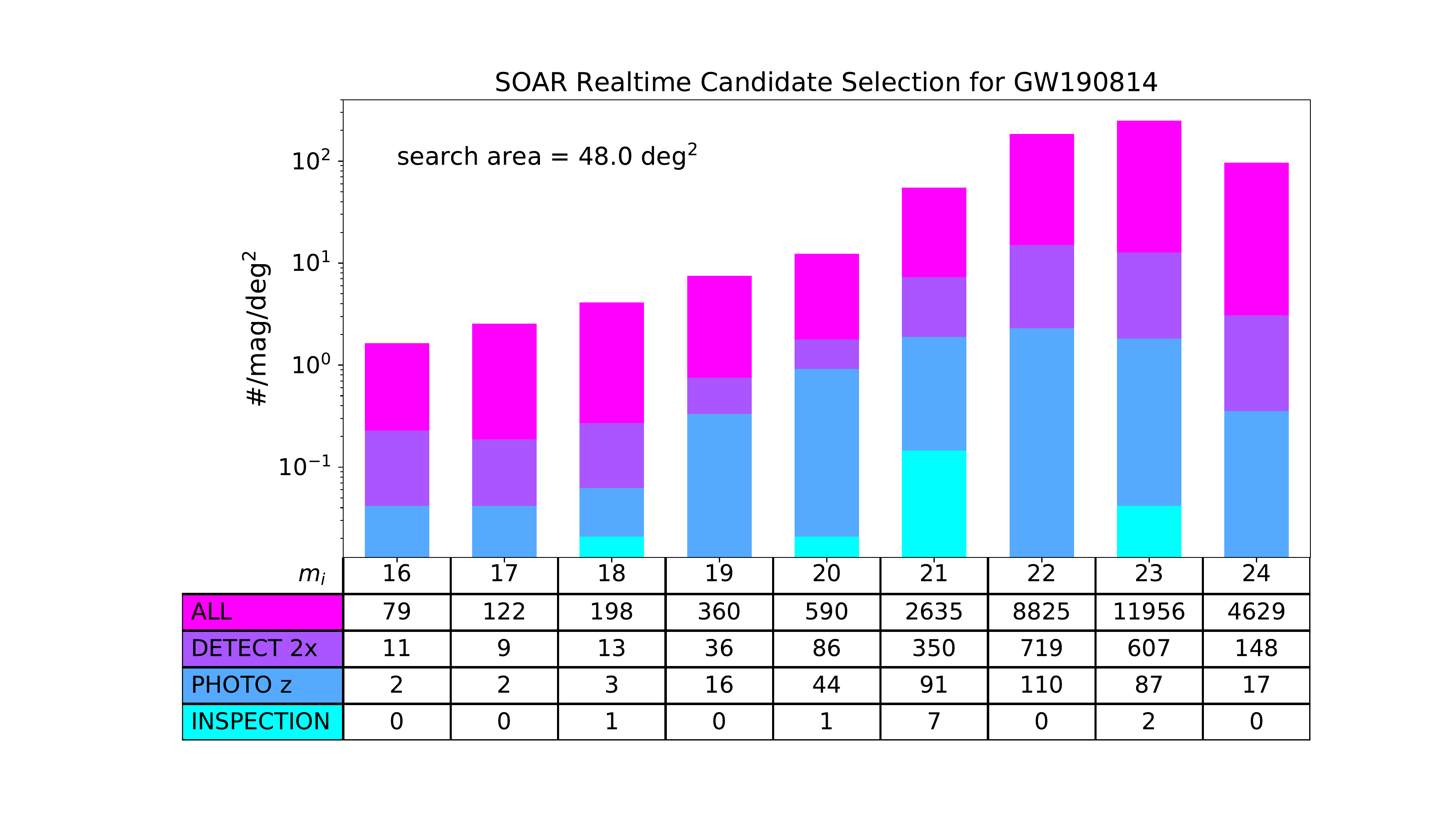}
\caption{The DECam search \& discovery candidate selection for spectroscopic follow-up for GW190814. 
Whereas Fig. \ref{fig:candidates_o3} provided the typical distribution of DECam candidates expected for a typical LVC O3 { BNS} merger, here we show the corresponding $i$-band magnitude distribution of all transient candidates observed and visually inspected and identified within the observed area by DECam across the selection criteria of \S \ref{sec:filter} specifically for the GW event GW190814. 
The final 11 candidates targeted with SOAR compose the cyan histogram and the { {\tt ``INSPECTION''}} row; 4 other candidates, which were in the $i=21-22$ range, were observed by other telescopes and are omitted from the cyan histogram and { {\tt ``INSPECTION''}} row. Note that at the time of SOAR follow-up on three of these transients, their magnitudes had faded below the SOAR detection limit, so we observed their host galaxies to measure their redshifts.  (Note:  the numbers listed below the plot are the total per magnitude bin for the full 48.0~sq~deg search area; the $y$-axis of the plot, however, is the number per magnitude bin {\em per square degree}.)
}
\label{fig:candidates_gw190814}
\end{figure}

Whether an object was first reported to the GCN by the DESGW team or by another follow-up team, it was still required to pass the same set of selection criteria prior to being targeted with SOAR.
Technical details 
and motivations for these criteria 
are presented in \citet{Morgan_paper}.
Remaining objects after the above selection criteria were sorted by their single-band average rate of change in flux to look for rapidly evolving transients.
Finally, we triggered SOAR on objects passing the criteria and that had not already been ruled out by other teams in order of largest flux change to smallest flux change\footnote{Those candidates ruled out by other teams included candidates observed on the The Gran Telescopio Canarias (GTC; \citealt{2019GCN.25419....1L, 2019GCN.25543....1C,2019GCN.25571....1L,2019GCN.25588....1H}), The Southern African Large Telescope (SALT; \citealt{Morgan_paper}), and The Giant Magellan Telescope (GMT; \citealt{Morgan_paper}), and in general were too faint for SOAR ToO follow-up.}.  The selection process for the specific case of 
GW190814 is illustrated in Figure~\ref{fig:candidates_gw190814}.

\begin{figure}
\centering
\includegraphics[width=1.0\linewidth]{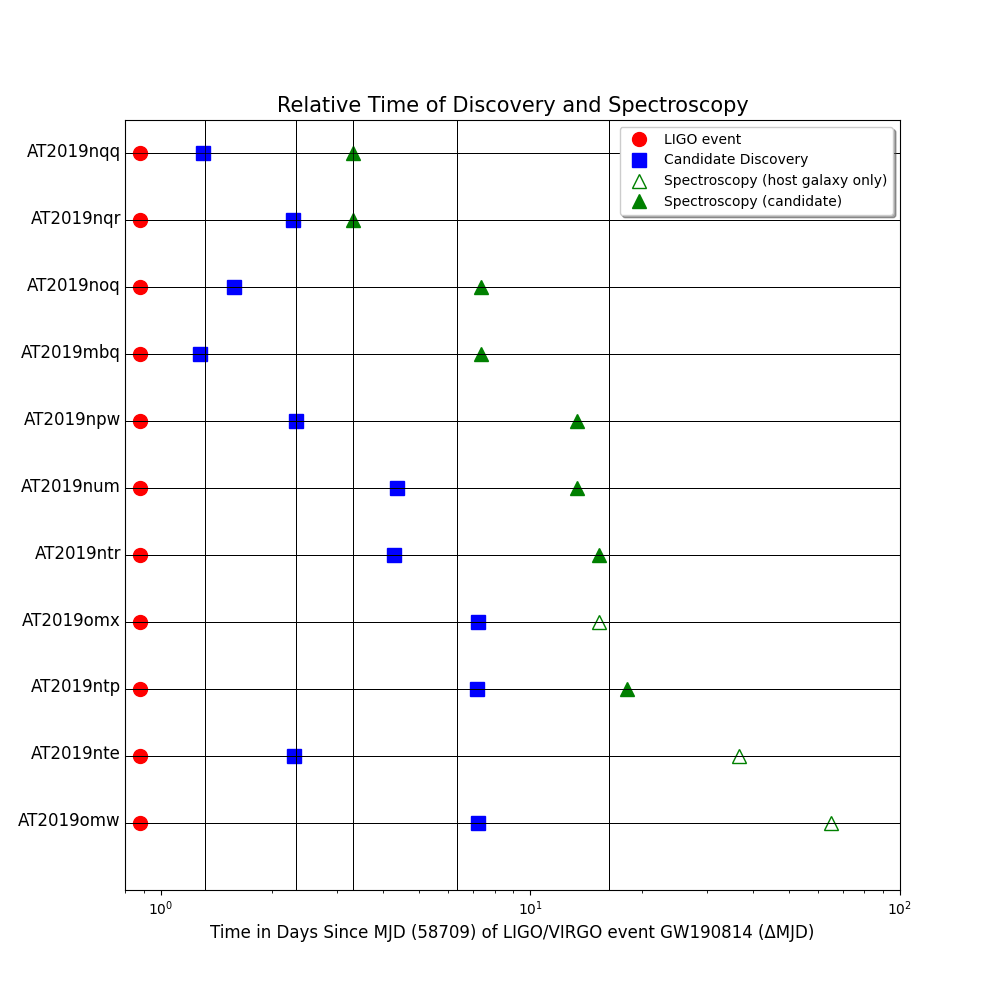}
\caption{{ Observational t}imelines for each 
{ KN} candidate. All dates are shown as number of days ($\Delta$MJD) since 58709.00, MJD corresponding to August 14, 2019, the day GW190814 { was detected}.  The { time of the NSBH merger event} at MJD 58709.88 is shown (using a red circle) on each. The date of transient discovery is shown as a blue square. The date of SOAR spectroscopy is shown as a green triangle for each 
{ KN} candidate (open triangles indicate that spectroscopy was only done for the host galaxy).  Vertical lines show beginning time of DECam observations.}
\label{fig:timeline}
\end{figure}

In total, 11 objects were targeted with SOAR for either spectroscopic classification of the transient or to obtain a spectroscopic redshift of the host-galaxy. 
These objects are cataloged in Table \ref{tab:soar_obs} and their times of photometric discovery and spectroscopic follow-up are shown visually in Figure \ref{fig:timeline}.  We note that the observed rate (11 candidates within 48~sq~deg) well matches the anticipated rate (9 candidates within 60~sq~deg), and are in fact identical within the Poisson errors.

\begin{figure*}
\centering
\vspace{-.75cm}
\includegraphics[width=1.0\textwidth]{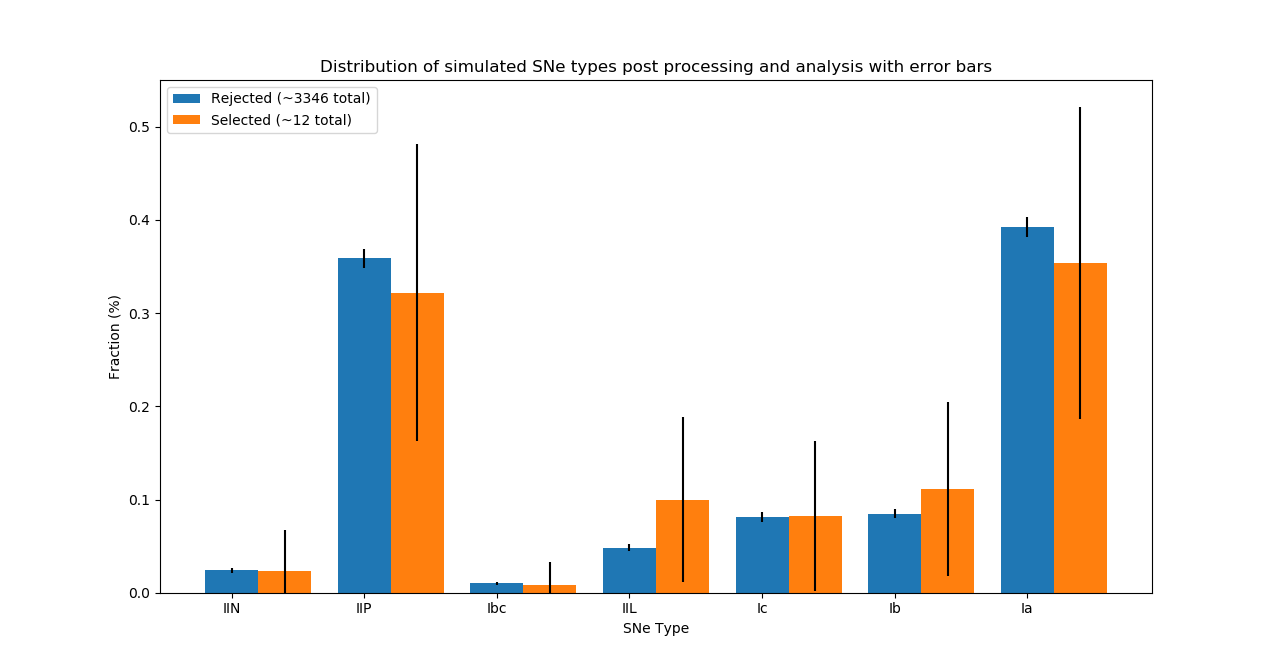}
\caption{Predictions of the relative incidence of each of several types of { SN} within a spectroscopic follow-up 
{ KN} candidate sample post DECam processing \& analysis.  The predictions are based on simulated data using SNANA light-curves and PLAsTiCC models and run through the selection steps of \citet{Morgan_paper}.  The blue histogram shows the relative distribution of { SNe} that were rejected by the selection steps; the orange histogram, the relative distribution of { SNe} that survived ({\em i.e.\/} were selected by) all the selection steps.  Similar relative sizes of bars indicates no bias towards any particular { SN} type. The error largely comes from the Poisson counting statistics. }
\label{fig:sim_results}
\end{figure*}

In Figure \ref{fig:sim_results} we show the expected incidence of each of several types of { SN} during a search for a 
{ KN}. These data come from simulated full light curves using the SuperNova ANAlysis software (SNANA; see \S~\ref{sec:software}). The models are the same as in the Photometric LSST Astronomical Time-series classification challenge (PLAsTiCC, \citealt{kessler2019}). We start with $\approx$3300 { SNe} with a distribution of { SN} types at random points in their light curves -- what one might net in a typical transient search by DECam covering several tens of square degrees -- and then apply the selection (culling) steps detailed 
above, 
in the end yielding about a dozen SNe whose imaging and photometric properties closely enough mimic that of a
{ KN} that they would require follow-up spectroscopy 
(and/or a more robust photometry-based technique) 
to eliminate them as candidates in a 
{ KN} search.  This could be viewed as an estimate of the rough contamination rate by { SNe} 
in a real-time imaging search using similar candidate selection criteria.
Finally, it is interesting to note that the distribution of { SN} types is very similar between the sample of 3346 { SNe} that were rejected by the 
above selection steps and the sample of a dozen { SNe} that successfully passed through all these steps.  In other words, the selection steps do not seem to favor or disfavor any particular SN type.

\section{SOAR Observations}
\label{sec:soar_obs}
In the following section (\S \ref{sec:SOAR_proc}) we provide details of our 
ToO triggers and real-time ({\em not\/} final) classifications in search of the optical counterpart of GW190814. 
We explain how the methods described in \S \ref{sec:spectro_candidate_select} were executed when our SOAR 2019B ToO program was triggered to observe candidates for an  optical counterpart of GW190814.

\subsection{GW190814 candidate observations}\label{sec:SOAR_proc}

Based on input from the DECam search \& discovery program, we developed a list of candidates for spectroscopy as described in the previous section. For the objects possible to observe each night we developed nightly webpages with information on object airmasses, finding charts and other information that would be required once our ToO time began. On each night we issued a ToO interrupt, there were several possible kilonova candidates that could be observed. The selection of which ones were to be targeted for the night was based on observing conditions (e.g.\ low airmass) and brightest magnitude.  

In order to complete data processing in real time, we employed a custom-made reduction pipeline that we developed, a Jupyter notebook we call the SOAR Goodman Quick Reduce ({ see} \S~\ref{sec:software}), to obtain quick results immediately after the data are transferred from the SOAR telescope machines. The preliminary processing consists of a quick reduction of the spectra using an arc-lamp wavelength calibration frame  
and a calibration from a standard star taken at the start of ToO observing. This { publicly available} Jupyter notebook takes the 2D spectrum, extracts the 1D spectrum, and performs basic wavelength and spectrophotometric calibration with relatively simple and straightforward inputs. With a little practice, it is time-competitive with just using the IRAF implot task -- but with the added advantage of providing a quick 
calibrated spectrum.   Generally, a ``by eye'' check of the calibrated spectrum indicates whether or not a  candidate is a 
{ KN} -- usually due to the disqualifying presence of one or more relatively sharp emission lines or the spectral features of { an SN} -- but, even so, each calibrated spectrum was also sent that same night to one of our { SN}-fitting experts, who would fit the spectrum to { SN} model spectra.
The resulting spectra were intended to be analyzed with fast classification tools (see below) and the spectroscopic class and redshift of the transient to be published promptly to the community via a GCN circular. 
The list of objects for which spectra were taken, along with initial redshift and { SN} classifications and the GCNs the DESGW SOAR observing team issued, is given in Table \ref{tab:soar_obs}.

{ To avoid fatigue, the DESGW SOAR spectroscopy task force was divided into four teams -- a team based in Brazil (PI M. Makler), a team based in Chile (PI F. Olivares), a team based at UC-Santa Cruz (PI C. Kilpatrick), and a team based at Fermilab (PI D. Tucker) -- each team signing up for multiple 2-week shifts throughout the course of LVC O3.}  
Our default plan was to use the Goodman HTS Blue camera, the 400~l/mm grating in its M1 configuration, and a slit width of 1~arcsec, to yield a wavelength range of roughly 3000\AA \ to 7050\AA\ at a resolution of $R \sim 930$ (e.g., see \citealt{nicholl17}), but, if the night's main program that our ToO was interrupting was using a roughly similar configuration, we could also use that instead, minimizing issues with switch-overs from and to the main program.  

\subsubsection{Observations}

We issued ToO interrupts on 2019 August 16, 20, 26, 28, and 31 (start dates, based on local time). On several other nights we attempted to conduct ToO observations, but found skies to be too cloudy to effectively observe and so we canceled the ToO interrupts.  During the course of the August 2019 observations, the Fermilab and Chilean teams were on shift.  In addition, spectra were taken for us by SOAR scientific staff during the SOAR engineering nights of September { 13} (host galaxy for AT2019nte) and October { 17} (host galaxy of AT2019omw). This information and the GCNs issued are summarized in Table \ref{tab:soar_obs}. 


\begin{deluxetable*}{llllll}
\tablehead{
\colhead{Candidate} & \colhead{Night} & \colhead{GCN} & \colhead{Classification Source} &
\colhead{Classification} & \colhead{Redshift} 
}
\startdata
AT2019nqq & Aug 16  & 25379 & Astrodash                      & Type Ic-broad SN                   & 0.3257  \\
AT2019nqr & Aug 16  & 25379 & Astrodash                      & Type IIb SN                        & 0.0888  \\
AT2019noq & Aug 20  & 25423 & SNID                           & Type IIP SN                        & 0.07  \\
AT2019mbq & Aug 20  & 25423 & SNID                           & Type Ia-CSM SN                     & 0.10 \\
AT2019npw & Aug 26  & 25484 & Astrodash                      & Type IIb SN                        & 0.163 \\
AT2019num & Aug 26  & 25484 & Astrodash                      & Type IIP SN                        & 0.113 \\
AT2019ntr & Aug 28  & 25540 & Astrodash                      & Type II-L SN                       & 0.2 \\
AT2019omx & Aug 28  & 25540 & H$\alpha$ emission line        & host galaxy                        & {0.275}$^\star$  \\
AT2019ntp & Aug 31  & 25596 & Astrodash                      & Type Ic-BL SN                      & 0.3284 \\
AT2019nte & Sep 13  & 25784 & H$\alpha$/[NII] emission lines & host galaxy                        & {0.0704}$^\star$  \\
AT2019omw & Oct 17  & N/A   & H$\alpha$ emission line        & host galaxy                        & {0.0467}$^\star$  \\
\enddata
\caption{Initially reported data for the 11 candidates described in this paper. Data include candidate name as assigned by the Transient Name Server, night of observation, GCN in which spectral results were reported, source of initial classification and redshift, initial classification and initial redshift. These are the values reported in the GCNs.  (No GCN was submitted for AT2019omw.) These values were updated after full reduction and processing of data. Updated values are given in Table \ref{tab:candidates3}. (Astrodash and SNID are { SN} spectrum fitting codes; see \S~\ref{sec:reliability} and \S~\ref{sec:software}.  Which fitting code was used in this initial classification for a given candidate depended heavily on which team member was available on that night to perform the classification, and the team member's preference.)  \label{tab:soar_obs}}
\tablenotetext{*}{Redshift of the host galaxy.}
\tablecomments{Night=civil date of the start of the night of observation, the NOAO convention of designating an observing night. The asterisk to the right of several $z$ values indicates that this is redshift for the host galaxy, as the transient was too dim to observe.}
\end{deluxetable*}

In Figure \ref{fig:timeline} we graphically summarize our sequence of observations.  In this figure we show a set of timelines indicating the dates of discovery and SOAR spectroscopy of each of the candidates we observed, using a log scale for the x-axis. The first mark (red circle) on each timeline is the MJD of the GW190814 { merger} event. The second mark (blue square) is the date of discovery in  DECam observations.  The third mark (green triangle) indicates the date of SOAR spectroscopy.  Vertical lines are also included that show the date of DECam observations, as described in \citet{Morgan_paper}. The marks denoting SOAR spectroscopy of AT2019nte, AT2019omw, and AT2019omx,  are unfilled, indicating that we did not take spectroscopy of the transient but of the host galaxy only. We report redshifts of these host galaxies in Table~\ref{tab:soar_obs}. The horizontal axis is given in $\Delta$MJD, time in days since MJD 58709.

\begin{deluxetable*}{l| r| l@{\,}l@{\,}l@{\,}l|l@{\,}l@{\,}l@{\,}l|l}
\tablecaption{Final results for the 8 transients and the 3 host galaxies for which we took spectra. Results include name from the Transient Name Server and the S/N of the spectrum calculated using the 6000-6100 $\AA$ region. Then we report the outputs from AstroDash and SNID, respectively, including { SN} type, $\tt{rlap}$ values, redshift, and absolute magnitude (at DECam  discovery; see Table~\ref{tab:candidates1}).  
For spectra with $S/N < 5$ and for fits with $\tt{rlap} < 6.0$ (AstroDash) or $\tt{rlap} < 5.0$ (SNID), the classification may be unreliable.\label{tab:candidates3}} 
\tablehead{
\colhead{}  & \colhead{} &\multicolumn{4}{c}{AstroDash}  & \multicolumn{4}{c}{SNID} & \colhead{Comments} \\
\colhead{Name / ID} & \colhead{$S/N$} & 
\colhead{Type} & \colhead{$\texttt{rlap}$} & \colhead{$z$} & \colhead{$M_{\text{abs}}$} & 
\colhead{Type} & \colhead{$\texttt{rlap}$} & \colhead{$z$} & \colhead{$M_{\text{abs}}$} & \colhead{} 
}
\startdata
%
%
AT2019nqq$^{\dagger}$     &  2.4 & Ia-csm   &  0.14 & 0.071 &~$-16.8$  & IIn      &  5.3  & 0.070 &~$-16.8$ & SNID preferred\\
AT2019nqr                 & 32.6 & Ia-csm   &  9.97 & 0.086 &~$-19.6$  & Ia       &  4.36  & 0.101 &~$-20.0$ & Seyfert 2 AGN @ $z=0.083$\\
AT2019noq                 &  7.7 & IIn      & 19.55 & 0.074 &~$-17.7$  & IIP      & 13.11 & 0.072 &~$-17.6$ & AstroDash preferred\\
AT2019mbq$^{\dagger}$     & 23.1 & IIn      & 15.96 & 0.102 &~$-17.6$  & Ia       & 12.09 & 0.110 &~$-17.8$ & AstroDash preferred\\
AT2019npw                 &  6.4 & IIP      &  4.76 & 0.148 &~$-18.7$  & IIP      &  6.44 & 0.148 &~$-18.7$ & SNID preferred\\
AT2019num$^{\dagger}$     &  7.5 & IIL      &  7.95 & 0.123 &~$-17.5$  & IIb      &  6.96 & 0.149 &~$-18.0$ & AstroDash preferred\\
AT2019ntr$^{\dagger}$     &  1.8 & Ic-broad &  0.81 & 0.224 &~$-19.0$  & Ia       &  4.01 & 0.861 &~$-22.5$ & None preferred; unknown\\
AT2019omx$^{* \dagger}$   &  2.3 &  ...     &  ...  &  ...  &  ...     &  ...     &  ...  &  ...  &  ...    & host galaxy @ $z=0.275$ ($M_{\rm abs}=-18.7$)\\
AT2019ntp                 & 11.8 & Ia-pec   &  6.44 & 0.116 &~$-17.7$  & Ia       & 12.22 & 0.114 &~$-17.6$ & SNID preferred\\
AT2019nte$^{* \dagger}$   &  5.8 &  ...     &  ...  &  ...  &  ...     &  ...     &  ...  &  ...  &  ...    & host galaxy @ $z=0.0704$ ($M_{\rm abs}=-16.6$)\\
AT2019omw$^{*}$           &  1.8 &  ...     &  ...  &  ...  &  ...     &  ...     &  ...  &  ...  &  ...    & host galaxy @ $z=0.0467$ ($M_{\rm abs}=-13.8$)\\
\enddata
\tablenotetext{*}{Only the spectrum of the host galaxy was obtained; so it was not fit by either AstroDash or SNID.}
\tablenotetext{\dagger}{This candidate lies outside the 90\% confidence probability contours of the final LVC map for GW190814; see Fig.~\ref{fig:prob_map}.}
\end{deluxetable*}
\begin{figure}
\centering
\vspace{-.75cm}
\hspace{-.7cm}\includegraphics[width=1.10\linewidth]{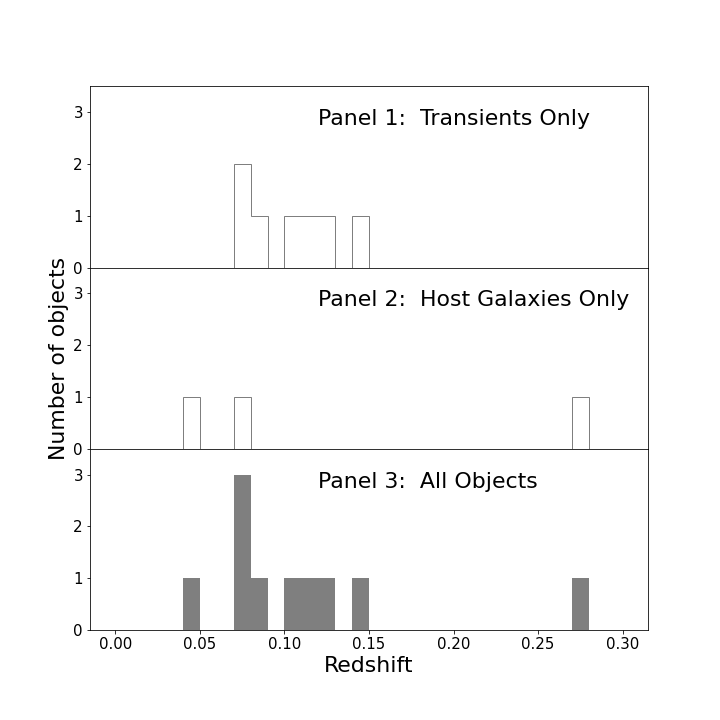}
\caption{Histograms of the redshifts of the eleven candidates, using final preferred results from Table \ref{tab:candidates3}. The top panel is for the 8 transient targets alone, the middle panel is for for the 3 host galaxy targets alone, and the bottom panel is for all 11 SOAR targets combined  (transients and host galaxies together).}
\label{fig:properties}
\end{figure}

Even though none of these 11 candidates were determined to be the optical counterpart of GW190814, these results will permit important upper limits to be established in preparation for future searches for the optical counterparts of these types of mergers (see next section).

\section{Results \& Discussion} 
\label{sec:results}
\begin{figure*}
\centering
\begin{tabular}{|c c|}
\hline
\multicolumn{2}{|c|}{} \\
\multicolumn{2}{|c|}{\textbf{\LARGE{AT2019noq}}} \\
\multicolumn{2}{|c|}{} \\
\textbf{Finding Chart} & \textbf{DECam light curve} \\
\includegraphics[width=0.4\textwidth]{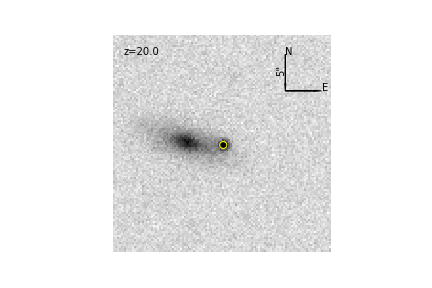} &
\includegraphics[width=0.4\textwidth]{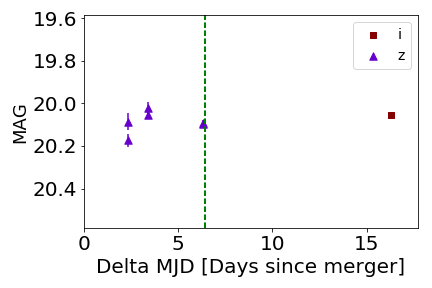} \\ 
\multicolumn{2}{|c|}{} \\
\textbf{Fit to SN spectrum templates} & \textbf{Fit to KN spectrum models} \\
\includegraphics[width=0.4\textwidth]{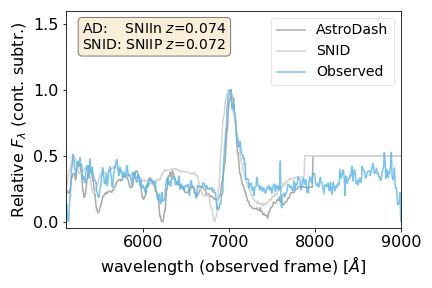} & \includegraphics[width=0.4\textwidth]{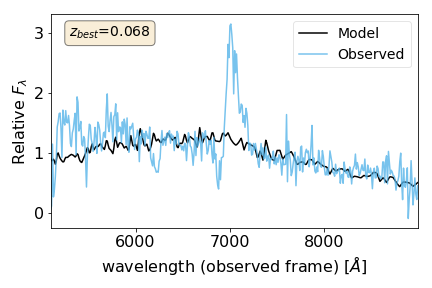} \\
\hline
\end{tabular}  
\caption{Top Left: The thumbnail finding chart (using the DECam imaging) for the AT2019noq  
{ KN} candidate; the location of the candidate is marked by a small yellow circle. Top Right: the candidate's $i$- and $z$-band light curves from DECam photometry; the vertical dashed green indicates when SOAR spectroscopy was obtained.  Bottom Left: Observed and best-fit SN model spectrum for the candidate object. Light blue is the processed, calibrated, and continuum-subtracted observed spectrum; dark grey is the best-fit { SN} model from AstroDash; and light grey is the best-fit { SN} model from SNID. In the panel we provide the best-fit SN type and redshift from the two codes.  Bottom Right:  Observed and best-fit model 
{ KN} spectra for the candidate objects. Light blue is the processed and calibrated observed spectrum; black is the best fit \citet{kasen_2017} 
{ KN} model. In the panel we provide the best-fit value of the redshift, $z_{\rm best}$. Unlike in AstroDash/SNID fits plot, the continuum has not been subtracted.  Also, a slightly different smoothing technique is used for the SN fits and for the KN fits.}
\label{fig:thumbnails-all4}
\end{figure*}

\begin{figure*}
\centering
\begin{tabular}{|c c|}
\hline
\multicolumn{2}{|c|}{} \\
\multicolumn{2}{|c|}{\textbf{\LARGE{AT2019mbq}}} \\
\multicolumn{2}{|c|}{} \\
\textbf{Finding Chart} & \textbf{DECam light curve} \\
\includegraphics[width=0.4\textwidth]{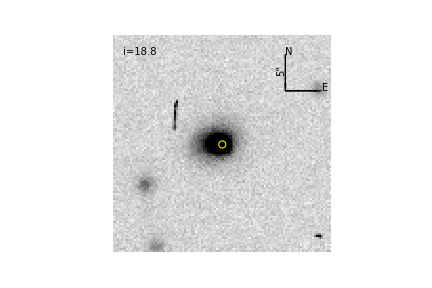} &
\includegraphics[width=0.4\textwidth]{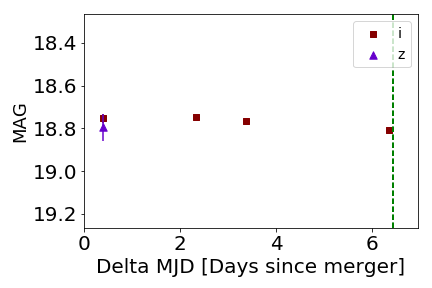} \\ \multicolumn{2}{|c|}{} \\
\textbf{Fit to SN spectrum templates} & \textbf{Fit to KN spectrum models} \\
\includegraphics[width=0.4\textwidth]{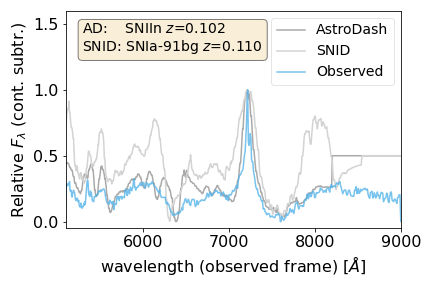} & \includegraphics[width=0.4\textwidth]{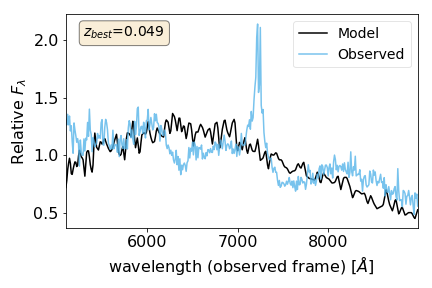} \\
\hline
\end{tabular}  
\caption{Same as Fig.~\ref{fig:thumbnails-all4} except for the
AT2019mbq 
{ KN} candidate.}
\label{fig:thumbnails-all5}
\end{figure*}

\begin{figure*}
\centering
\begin{tabular}{|c c|}
\hline
\multicolumn{2}{|c|}{} \\
\multicolumn{2}{|c|}{\textbf{\LARGE{AT2019npw}}} \\
\multicolumn{2}{|c|}{} \\
\textbf{Finding Chart} & \textbf{DECam light curve} \\
\includegraphics[width=0.4\textwidth]{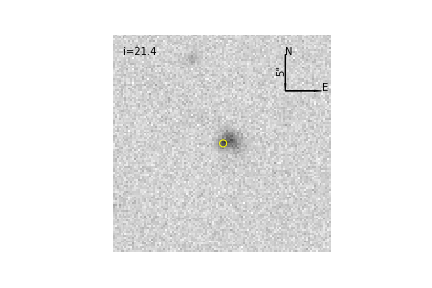} &
\includegraphics[width=0.4\textwidth]{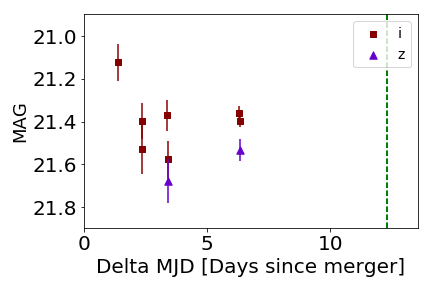} \\ \multicolumn{2}{|c|}{} \\
\textbf{Fit to SN spectrum templates} & \textbf{Fit to KN spectrum models} \\
\includegraphics[width=0.4\textwidth]{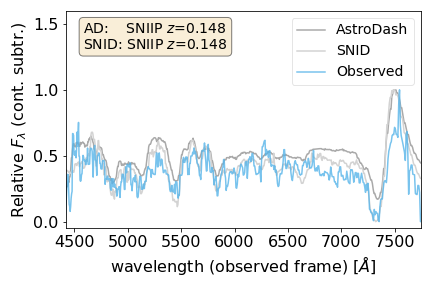} & \includegraphics[width=0.4\textwidth]{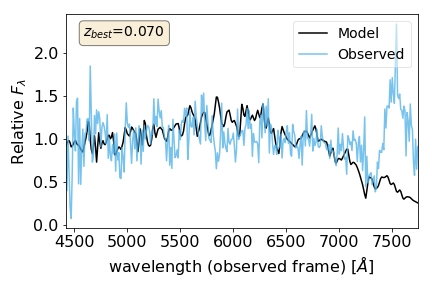} \\
\hline
\end{tabular}  
\caption{Same as Fig.~\ref{fig:thumbnails-all4} except for the AT2019npw 
{ KN} candidate.}
\label{fig:thumbnails-all6}
\end{figure*}

\begin{figure*}
\centering
\begin{tabular}{|c c|}
\hline
\multicolumn{2}{|c|}{} \\
\multicolumn{2}{|c|}{\textbf{\LARGE{AT2019num}}} \\
\multicolumn{2}{|c|}{} \\
\textbf{Finding Chart} & \textbf{DECam light curve} \\
\includegraphics[width=0.4\textwidth]{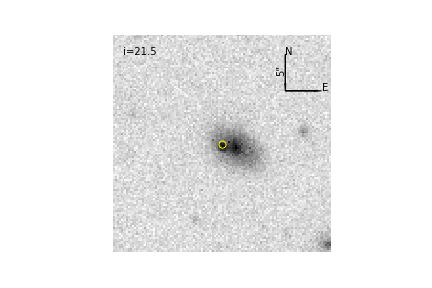} &
\includegraphics[width=0.4\textwidth]{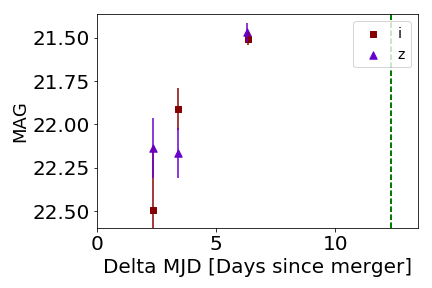} \\ \multicolumn{2}{|c|}{} \\
\textbf{Fit to SN spectrum templates} & \textbf{Fit to KN spectrum models} \\
\includegraphics[width=0.4\textwidth]{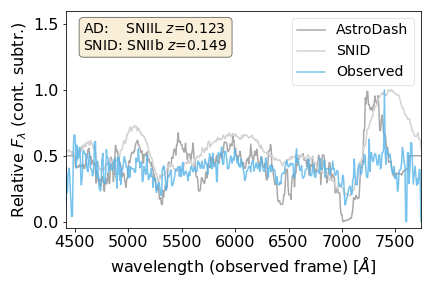} & \includegraphics[width=0.4\textwidth]{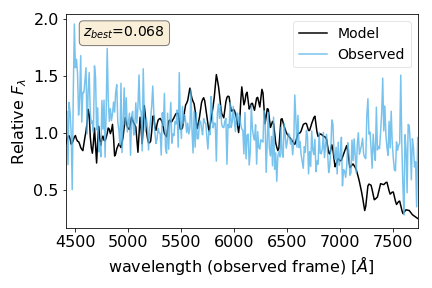} \\
\hline
\end{tabular}  
\caption{Same as Fig.~\ref{fig:thumbnails-all4} except for the AT2019num 
{ KN} candidate.}
\label{fig:thumbnails-all7}
\end{figure*}

\begin{figure*}
\centering
\begin{tabular}{|c c|}
\hline
\multicolumn{2}{|c|}{} \\
\multicolumn{2}{|c|}{\textbf{\LARGE{AT2019ntr}}} \\
\multicolumn{2}{|c|}{} \\
\textbf{Finding Chart} & \textbf{DECam light curve} \\
\includegraphics[width=0.4\textwidth]{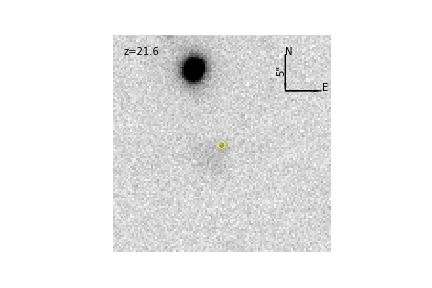} &
\includegraphics[width=0.4\textwidth]{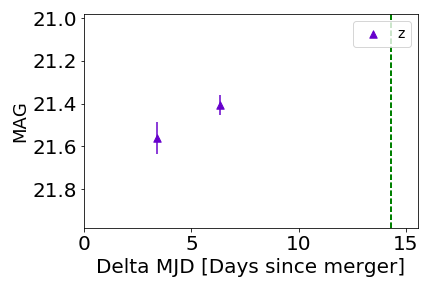} \\ 
\multicolumn{2}{|c|}{} \\
\textbf{Fit to SN spectrum templates} & \textbf{Fit to KN spectrum models} \\
\includegraphics[width=0.4\textwidth]{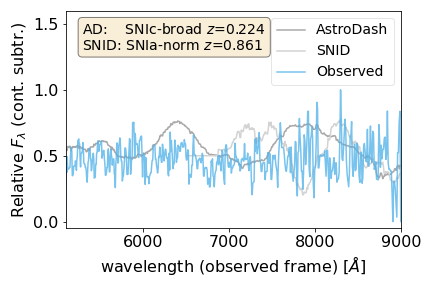} & 
\includegraphics[width=0.4\textwidth]{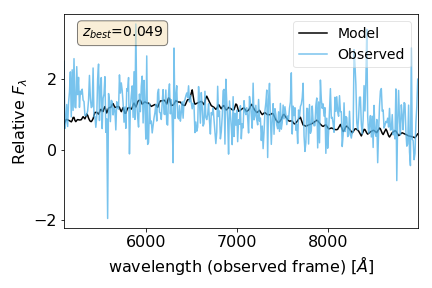} \\
\hline
\end{tabular}  
\caption{Same as Fig.~\ref{fig:thumbnails-all4} except for the AT2019ntr 
{ KN} candidate.}
\label{fig:thumbnails-all8}
\end{figure*}

\begin{figure*}
\centering
\begin{tabular}{|c c|}
\hline
\multicolumn{2}{|c|}{} \\
\multicolumn{2}{|c|}{\textbf{\LARGE{AT2019ntp}}} \\
\multicolumn{2}{|c|}{} \\
\textbf{Finding Chart} & \textbf{DECam light curve} \\
\includegraphics[width=0.4\textwidth]{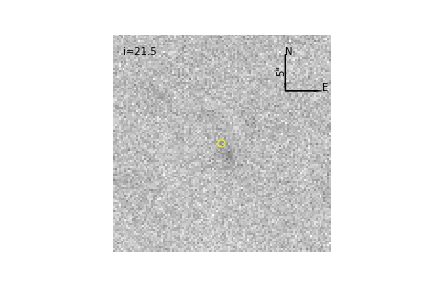} &
\includegraphics[width=0.4\textwidth]{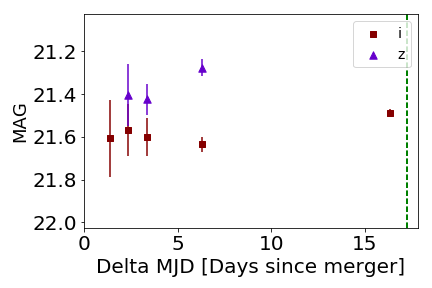} \\ 
\multicolumn{2}{|c|}{} \\
\textbf{Fit to SN spectrum templates} & \textbf{Fit to KN spectrum models} \\
\includegraphics[width=0.4\textwidth]{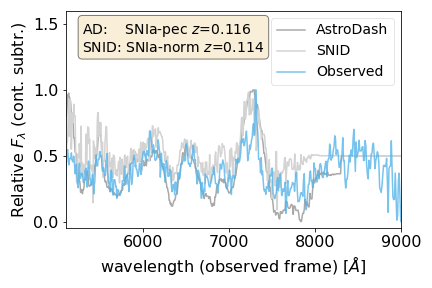} & 
\includegraphics[width=0.4\textwidth]{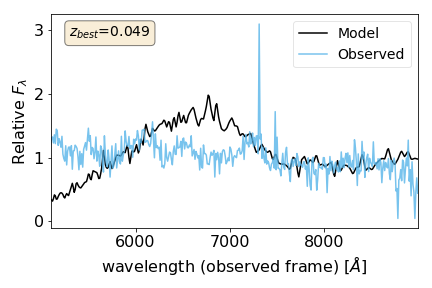} \\
\hline
\end{tabular}  
\caption{Same as Fig.~\ref{fig:thumbnails-all4} except for the AT2019ntp 
{ KN} candidate.  (Due to the additional smoothing in the SN-fitting plot, the strong narrow emission line seen in the KN-fitting plot is mostly washed out.)}
\label{fig:thumbnails-all9}
\end{figure*}

\begin{figure*}
\centering
\begin{tabular}{|c c|}
\hline
\multicolumn{2}{|c|}{} \\
\multicolumn{2}{|c|}{\textbf{\LARGE{AT2019nqr}}} \\
\multicolumn{2}{|c|}{} \\
\textbf{Finding Chart} & \textbf{DECam light curve} \\
\includegraphics[width=0.4\textwidth]{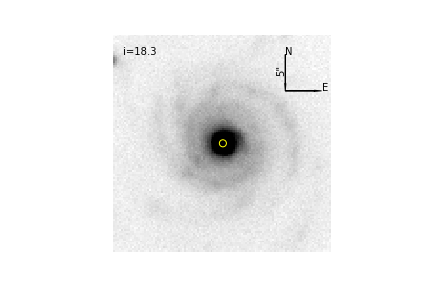} &
\includegraphics[width=0.4\textwidth]{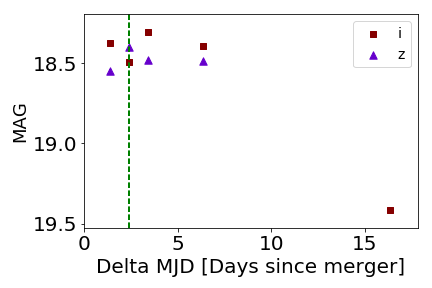} \\
\multicolumn{2}{|c|}{} \\
\textbf{Fit to SN spectrum templates} & \textbf{Fit to KN spectrum models} \\
\includegraphics[width=0.4\textwidth]{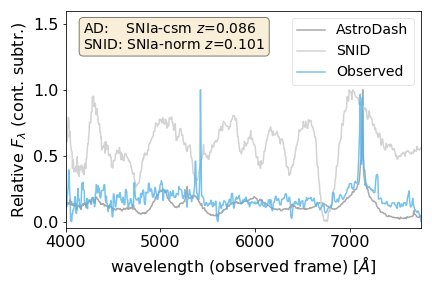} &
\includegraphics[width=0.4\textwidth]{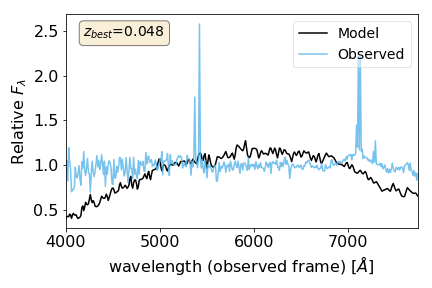}  \\
\multicolumn{2}{|c|}{} \\
\textbf{Fit to an AGN spectrum template} &  \\
\includegraphics[width=0.4\textwidth]{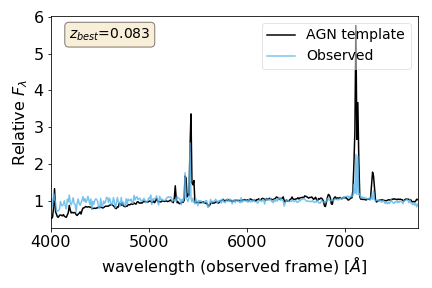} & \\
\hline
\end{tabular}  
\caption{Same as Fig.~\ref{fig:thumbnails-all4} except for the AT2019nqr 
{ KN} candidate. We also show the best fit to AGN template spectra, which is that of a Seyfert~2.}
\label{fig:thumbnails-all10}
\end{figure*}
\begin{figure*}
\centering
\begin{tabular}{|c c|}
\hline
\multicolumn{2}{|c|}{} \\
\multicolumn{2}{|c|}{\textbf{\LARGE{AT2019nqq}}} \\
\multicolumn{2}{|c|}{} \\
\textbf{Finding Chart} & \textbf{DECam light curve} \\
\includegraphics[width=0.4\textwidth]{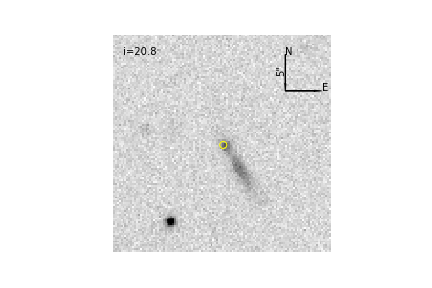} &
\includegraphics[width=0.4\textwidth]{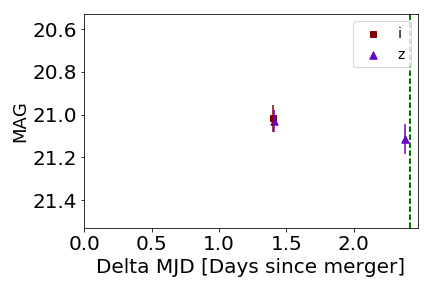} \\
\multicolumn{2}{|c|}{} \\
\textbf{Fit to SN spectrum templates} & \textbf{Fit to KN spectrum models} \\
\includegraphics[width=0.4\textwidth]{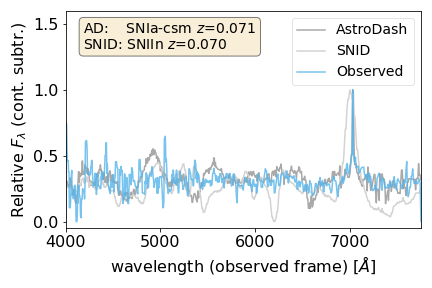} &
\includegraphics[width=0.4\textwidth]{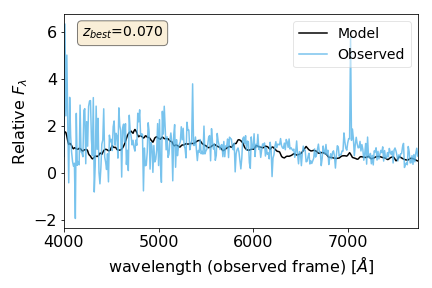} \\
\multicolumn{2}{|c|}{} \\
\textbf{Fit to an AGN spectrum template} &  \\
\includegraphics[width=0.4\textwidth]{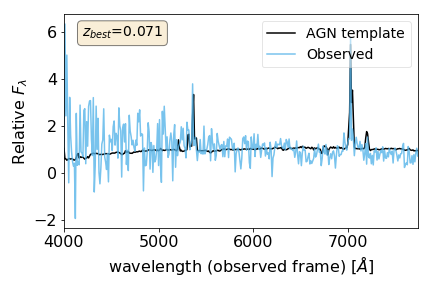} & \\
\hline
\end{tabular}  
\caption{Same as Fig.~\ref{fig:thumbnails-all4} except for the AT2019nqq 
{ KN} candidate. We also show the best fit to AGN template spectra, which is that of a Seyfert~2.}
\label{fig:thumbnails-all11}
\end{figure*}

\begin{figure*}
\centering
\begin{tabular}{|c c|}
\hline
\multicolumn{2}{|c|}{} \\
\multicolumn{2}{|c|}{\textbf{\LARGE{AT2019omx}}} \\
\multicolumn{2}{|c|}{} \\
\textbf{Finding Chart} & \textbf{DECam light curve} \\
\includegraphics[width=0.4\textwidth]{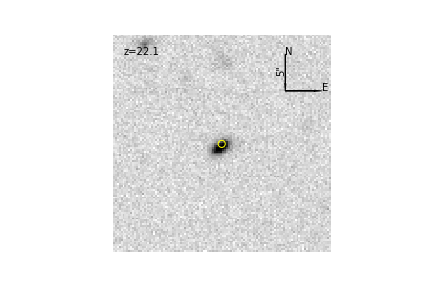} &
\includegraphics[width=0.4\textwidth]{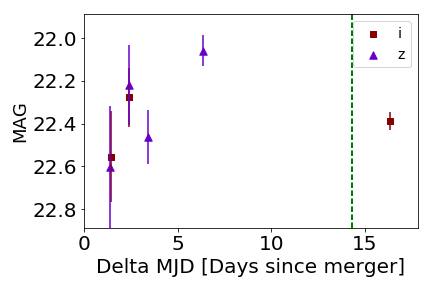} \\
\multicolumn{2}{|c|}{} \\
\textbf{Host galaxy spectrum} &  \\
\includegraphics[width=0.4\textwidth]{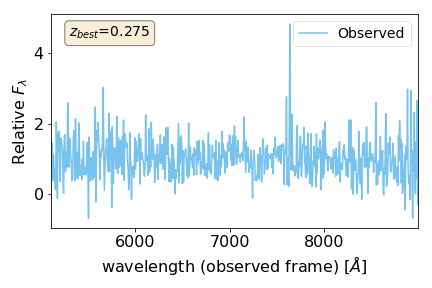} & \\
\hline
\end{tabular}  
\caption{
Top Left and Top Right: Same as Fig.~\ref{fig:thumbnails-all4} except for the AT2019omx 
{ KN} candidate. 
Bottom Left : The spectrum of the host galaxy.}
\label{fig:thumbnails-all1}
\end{figure*}

\begin{figure*}
\centering
\begin{tabular}{|c c|}
\hline
\multicolumn{2}{|c|}{} \\
\multicolumn{2}{|c|}{\textbf{\LARGE{AT2019nte}}} \\
\multicolumn{2}{|c|}{} \\
\textbf{Finding Chart} & \textbf{DECam light curve} \\
\includegraphics[width=0.4\textwidth]{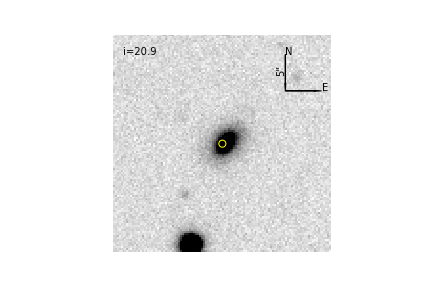} &
\includegraphics[width=0.4\textwidth]{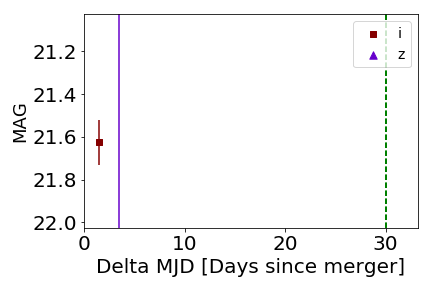} \\
\multicolumn{2}{|c|}{} \\
\textbf{Host galaxy spectrum} &  \\
\includegraphics[width=0.4\textwidth]{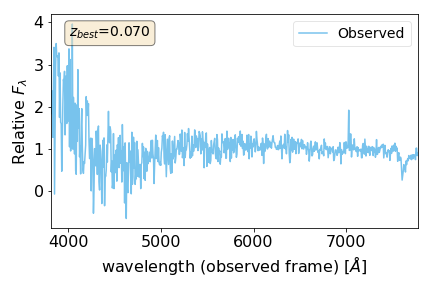} & \\
\hline
\end{tabular} 
\caption{Same as Fig.~\ref{fig:thumbnails-all1} but for the AT2019nte 
{ KN} candidate.  (The vertical purple line in the light-curve plot is just a very large error bar for the $z$-band observation.)}
\label{fig:thumbnails-all2}
\end{figure*}

\begin{figure*}
\centering
\begin{tabular}{|c c|}
\hline
\multicolumn{2}{|c|}{} \\
\multicolumn{2}{|c|}{\textbf{\LARGE{AT2019omw}}} \\
\multicolumn{2}{|c|}{} \\
\textbf{Finding Chart} & \textbf{DECam light curve} \\
\includegraphics[width=0.4\textwidth]{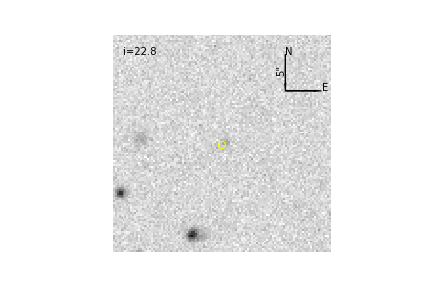} &
\includegraphics[width=0.4\textwidth]{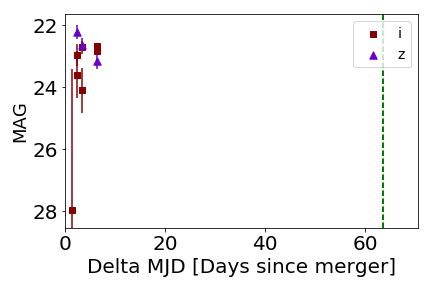} \\ 
\multicolumn{2}{|c|}{} \\
\textbf{Host galaxy spectrum} &  \\
\includegraphics[width=0.4\textwidth]{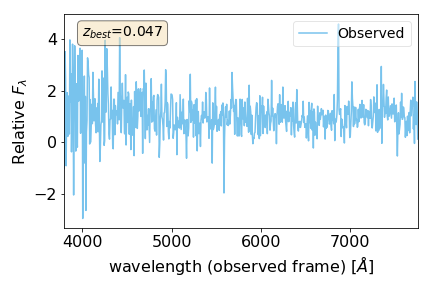} & \\
\hline
\end{tabular} 
\caption{Same as Fig.~\ref{fig:thumbnails-all1} except for the AT2019omw 
{ KN} candidate.}
\label{fig:thumbnails-all3}
\end{figure*}

In this section, we cover our final results from our SOAR observations of the GW190814 candidates. In \S~\ref{sec:spectral} we describe the full reduction and analysis of spectra and present the spectra themselves. In \S~\ref{sec:reliability} we present classifications of the supernovae and consider our methods of analysis. In \S~\ref{sec:Kasen} we fit each spectrum with \citet{kasen_2017} 
{ KN} models; as nearly all were found to be { an SN,} the 
{ KN} models are generally not good fits.  In \S~\ref{sec:host_galaxies}, we discuss the 3 candidates for which we only obtained spectra for the host galaxy and { the likelihood that either of these 3 candidates could be the optical counterpart for GW190814.}  Finally, in \S~\ref{sec:lessons} we consider lessons learned in LVC O3 that can be applied as we prepare for LVC observing season O4.

\subsection{Spectral data from SOAR Telescope}
\label{sec:spectral}

For the final reduced spectra (shown in Figs.\ \ref{fig:thumbnails-all4}
-- \ref{fig:thumbnails-all3}) --- unless otherwise noted\footnote{For the final reduced spectra for the host galaxies of AT2019nte and AT2019omw, we made use of standard IRAF reductions provided by the SOAR science staff.} --- we employed the UCSC spectral pipeline (link to Github repository in \S~\ref{sec:software}). This pipeline consists of the standard steps for the processing of optical spectroscopic data: bias subtraction, flat fielding, extraction of the 1D spectrum and flux and wavelength calibration against a standard star, typically a Hamuy Tertiary Standard Star \citep{Hamuy92, Hamuy94}. These more careful reductions, performed later, are the same as those used in the recent GW190914 omnibus paper by \citet{Kilpatrick_2021}.

\subsection{{ SN} Classifications}
\label{sec:reliability}

Offline analysis of the spectra we obtained was performed using the public codes Super Nova IDentification \citep[SNID;][]{SNID} and Deep Automated Supernova and Host classifier \citep[DASH, {\em a.k.a.\/}, AstroDash;][]{muthukrishna2019dash} (see \S~\ref{sec:software}). SNID is a template fitting method based on the correlation techniques by \citet{Tonry79}. 
 AstroDash is a deep convolutional neural network used to train a matching algorithm. These analysis tools provide spectral 
matching, which allowed us to classify our spectra by means of a comparison against a spectral library of transients and other astrophysical sources. We chose these codes as SNID has been used extensively 
{ by} the community and AstroDash makes use of a powerful deep learning technique. We discuss below the importance of using more than one { SN} typing package to check results.

For our AstroDash fits of the spectrum of each candidate, we applied an AstroDash smoothing length of 3 (unless otherwise stated), and we left the redshift a free parameter.  We then visually inspected the 20 best SN template fits for that candidate, choosing the top two for further consideration.  (The top two fits based on visual inspection also typically had among the highest  $\texttt{rlap}$ values of the 20 best fits.\footnote{$\texttt{rlap}$ is a measure of the quality of the fit that combines the correlation between the observed and the template spectrum with the amount of overlap in $\ln{\lambda}$-space between the observed and the template spectrum.  The higher the value of $\texttt{rlap}$, the higher the quality of the fit.  For the detailed definition, see \citet{SNID}.})  Unless there were other relevant considerations (e.g., the putative epoch in the light curve at which the spectrum was obtained), the SN template spectrum with the higher of the two $\texttt{rlap}$ values was chosen as the final best fit.

For our SNID fits of the spectrum of each candidate, we applied the default SNID smoothing length of 1 pixel, and, as with our AstroDash fits, we also fit for the redshift.  We visually inspected the top 5 SN template fits for each candidate, but in the end chose the one with the highest $\texttt{rlap}$ as our SNID classification.

In Table \ref{tab:candidates3} we present final measurements from AstroDash and from SNID for the 8 transients of which we took spectra. (For completeness, we also include information on the 3 candidates for which we only obtained host galaxy spectra:  AT2019omx, AT2019nte, and AT2019omw).  These results are based on the final reduced spectra. This table includes classification, the redshift, and a measure of the goodness of fit ($\tt{rlap}$) from these two { SN} spectrum fitting codes. We kept redshifts as free parameters in the fitting; the photometric redshifts of the host galaxies were used during the selection process of candidate objects discussed in \S~\ref{sec:decam_search}. 

The distribution of the redshifts from the preferred fits in Table~\ref{tab:candidates3} is given in Figure~\ref{fig:properties}; as expected, transients were found over a range of redshifts with a predominance of lower-$z$ objects. 

In Figures \ref{fig:thumbnails-all4}--\ref{fig:thumbnails-all3}, we provide the following information for each candidate:  a thumbnail finding chart containing the host galaxy and marking the location of the transient; the DECam-based $i$- and $z$-band light curves for the transients; and the final reduced observed spectrum.  For the candidates for which we only obtained the host galaxy spectrum,\footnote{Note that, within the 2.5~hour time constraint of a SOAR ToO interrupt, we were basically confined to observing targets that were $i\lesssim$21.5; so, in some cases -- especially for the later targets -- we instead obtained spectra of the candidate's host galaxy as a means of excluding the target by its redshift:  {\em i.e.\/}, if the redshift of the candidate's host galaxy is substantially discrepant from the redshift expected for the luminosity distance of the GW event ($z_{\rm GW} = 0.059 \pm 0.011$), we can exclude that candidate.} that is the sum of what we show in these figures.  For candidates for which we took a spectrum of the transient candidate itself, we also include the best-fit SN templates from AstroDash and SNID and the best-fit KN model from \citet{kasen_2017} overplotted on the final reduced observed spectrum.  As shown below, the interplay of these different types of data often helped in the final classification of a given candidate.

\subsubsection{AT2019noq}

For AstroDash, our two best fits were a $z=0.074$ SN~IIn 42--46 days past maximum light  ($\texttt{rlap}=19.55$) and a $z=0.079$ SN~IIP 2-6 days past maximum light ($\texttt{rlap}=19.31$).  The DECam light curve was relatively flat over the period it was observed (Fig.~\ref{fig:thumbnails-all4}); so we chose the SN~IIn classification as more likely.  For SNID, our best fit was a $z=0.072$ SN~IIP 9.8 days past maximum light  ($\texttt{rlap}=13.11$).  Due to its higher $\texttt{rlap}$ value, the AstroDash fit is preferred; see Figure~\ref{fig:thumbnails-all4}.

\subsubsection{AT2019mbq}

Recall that a spectrum of AT2019mbq was mistakenly observed by SOAR (the original target was AT2019ntn), and that there was evidence of a detection of AT2019mbq {\em before\/} the GW190814 merger event, making it highly unlikely that AT2019mbq is the optical counterpart.

For AstroDash, our two best fits were a $z=0.102$ SN~IIn 46--50 days past maximum light  ($\texttt{rlap}=15.96$) and a $z=0.103$ SN~IIn 42--46 days past maximum light ($\texttt{rlap}=14.92$).  The difference between the two classifications was small, and the DECam light curve provided no strong motivation to choose one over the  other (Fig.~\ref{fig:thumbnails-all5}); so we chose the template with the higher $\texttt{rlap}$ (a $z=0.102$ SN~IIn 46--50 days past maximum light) as more likely.  For SNID, our best fit was a 
$z=0.110$ 
SN~Ia 45.9 days past maximum light  
($\texttt{rlap}=12.09$).  Despite the SNID fit's relatively high $\texttt{rlap}$ value, a visual inspection of both the AstroDash and the SNID spectral fits (Fig.~\ref{fig:thumbnails-all5}) leads us to prefer the AstroDash fit.

\subsubsection{AT2019npw}

For AstroDash, our two best fits were a $z=0.148$ SN~IIP 18--22 days past maximum light  ($\texttt{rlap}=4.76$) and a $z=0.147$ SN~IIP 22--26 days past maximum light ($\texttt{rlap}=4.72$).  The difference between the two classifications was small, and the DECam light curve provided no strong motivation to choose one over the other; so we chose the template with the higher $\texttt{rlap}$ (a $z=0.148$ SN~IIP 18--22 days past maximum light) as more likely.  The relatively low $\texttt{rlap}$ values ($\texttt{rlap}< 6$), however, are of some concern.  For SNID, our best fit was a $z=0.148$ SN~IIP 44.3 days past maximum light  ($\texttt{rlap}=6.44$).  Due to its higher $\texttt{rlap}$ value, the SNID fit is preferred,
see Fig.~\ref{fig:thumbnails-all6}.

\subsubsection{AT2019num}

For AstroDash, our two best fits were a $z=0.123$ SN~IIL 6--10 days past maximum light  ($\texttt{rlap}=7.95$) and a $z=0.239$ SN~Ibn 22--26 days past maximum light ($\texttt{rlap}=0.4$).  Since the DECam light curve for this candidate is rising noticeably 10--6 days before the SOAR spectrum was obtained (Fig.~\ref{fig:thumbnails-all7}), it appears that this candidate is a likely a young SN; that, combined with the substantial difference in $\texttt{rlap}$ values led us to choose the $z=0.123$ SN~IIL 6--10 days past maximum light template as the more likely classification.  (We note that, for AT2019num, we used a smoothing length of 6 instead of 3 for our AstroDash fits.)  SNID, our best fit was a $z=0.149$ SN~IIb, 17.3 days {\em before\/} maximum light  ($\texttt{rlap}=6.96$).  Due to its higher $\texttt{rlap}$ value (and the relative rarity of catching a SN so early before maximum light), the AstroDash fit is preferred; see Figure~\ref{fig:thumbnails-all7}.

\subsubsection{AT2019ntr}

For AstroDash, our two best fits were a $z=0.224$ SN~Ic-broad near maximum light (between 2 days before and 2 days after peak; $\texttt{rlap}=0.81$) and a $z=0.264$ SN~Ia-csm 6--10 days past maximum light ($\texttt{rlap}=0.76$).  The DECam light curve seems to be slightly rising 11--8 days before the SOAR spectrum was taken (Fig.~\ref{fig:thumbnails-all8}), indicating a relatively young SN.  Due to the low $S/N$ of the spectrum (1.8) and the poor $\texttt{rlap}$ values for the fits, we are reluctant to assign a classification based on the AstroDash fits; that said, the $z=0.224$ SN~Ic-broad template near maximum light appears to be marginally better.

For SNID, our best fit was a $z=0.861$ SN~Ia 11.2 days {\em before\/} maximum light  ($\texttt{rlap}=4.01$).  Given a discovery $z$-band magnitude of $21.2$ (Table~\ref{tab:candidates1}), a redshift of $z=0.861$ implies a $z$-band absolute magnitude of roughly $M_{\rm abs}=-22.5$, or substantially more luminous than a typical SN~Ia \citep{2014AJ....147..118R}.  We therefore view the SNID fit as unreliable.

Due to the noisiness of this spectrum and the problems with both the AstroDash and the SNID fits, we prefer neither the AstroDash nor the SNID classifications.  We therefore view AT2019ntr's spectral classification as unknown; see Figure~\ref{fig:thumbnails-all8}.  {\em In hindsight, AT2019ntr would have been a natural candidate for additional spectroscopy with a larger telescope.\/}

\subsubsection{AT2019ntp}

For AstroDash, our two best fits were a $z=0.116$ SN~Ia-pec 34--38 days past maximum light ($\texttt{rlap}=6.44$) and a $z=0.331$ SN~Ic-Broad 26--30 days past maximum light  ($\texttt{rlap}=4.35$).  The DECam light curve provided no strong motivation to choose one over the  other (Fig.~\ref{fig:thumbnails-all9}); so we chose the template with the higher $\texttt{rlap}$ (a $z=0.116$ SN~Ia-pec 34--38 days past maximum light) as more likely.  For SNID, our best fit was a $z=0.114$ SN~Ia 45.8 days past maximum light  ($\texttt{rlap}=12.22$).  Due to its higher $\texttt{rlap}$ value, the SNID fit is preferred; see Figure~\ref{fig:thumbnails-all9}.

\subsubsection{AT2019nqr}
\label{sec:AT2019nqr_offline}

For AstroDash, our two best fits were a $z=0.086$ SN~Ia-csm 46-50 days past maximum light ($\texttt{rlap}=9.97$) and a $z=0.086$ SN~IIn 46--50 days past maximum light  ($\texttt{rlap}=7.85$).  We chose the template with the higher $\texttt{rlap}$ value as the better fit, despite that none of the SN templates did a reasonable job at fitting the narrow-but-strong emission lines at the observed wavelengths of 5371\AA ~and 5422\AA, and despite that the DECam light curve indicated that the transient may have been near a maximum brightness when the spectrum was observed.  For SNID, our best fit was a $z=0.101$ SN~Ia 5.7 days past maximum light ($\texttt{rlap}=4.36$). 
In the end, due to this candidate's central location in a spiral galaxy and a spectrum that well fits that of a Seyfert 2 at $z=0.083$, we classify AT2019nqr as a Seyfert 2 AGN; see Figure~\ref{fig:thumbnails-all10}.

\subsubsection{AT2019nqq}

For AstroDash, our two best fits were a $z=0.071$ SN~IIn 14--10 days {\em before\/} maximum light ($\texttt{rlap}=0.57$) and a $z=0.071$ SN~Ia-csm 6--10 days {\em past\/} maximum light  ($\texttt{rlap}=0.14$).  The DECam light curve appears to show a very slight fading over the short time it was monitored before the spectrum was taken (about 1 day before SOAR spectrum was obtained; Fig.~\ref{fig:thumbnails-all10}); so we chose the second template (a $z=0.071$ SN~Ia-csm 6--10 days past maximum light) as more likely, even though it has a lower $\texttt{rlap}$.   We note that the observed spectrum contains a prominent H$\alpha$ emission line redshifted to 7028\AA ~and a less prominent [O III]~5007 emission line redshifted to 5362\AA, and an even less prominent H$\beta$ emission line redshifted to 5205\AA.  For SNID, our best fit was a $z=0.070$ SN~IIn 50.2 days past maximum light  ($\texttt{rlap}=5.3$).  Due to its higher $\texttt{rlap}$ value, the SNID fit is preferred; see 
Figure.~\ref{fig:thumbnails-all11}.

We note that AT2019nqq was one system for which we could compare results from another facility. It was also observed by the GTC 10.4m (GCN25419), classified as a Type IIP { SN} at 4 days post maximum with $z_{host}$=0.071.  Although the type classification differs from our result for this system (Type IIn SN), the redshift estimate is consistent with ours.

\vspace{1.0cm}
In closing, we found that some classifications from both AstroDash and SNID might be inconclusive. For one case, AT2019ntr, this is probably related to the { low-$S/N$} spectrum, in which the low value of $\texttt{rlap}$ from both SNID and AstroDash points towards a poor fit.  It is also worth re-iterating that our methods of choosing the best fits differed for the two packages:  for AstroDash, we depended more on a visual inspection of the 20 models with the highest $\texttt{rlap}$ values; for SNID, we basically chose the model with the highest $\texttt{rlap}$ value.
This can lead to different classifications for the same object.  
In general, for a fit of a relatively high $S/N$ spectrum ($S/N \ge 5)$ and a relatively high value for $\texttt{rlap}$ ($\ge6.0$ for AstroDash; $\ge5.0$ for SNID), we view the classification (AstroDash or SNID) with the higher the value of $\texttt{rlap}$ as the preferred classification; in cases of a low $S/N$ spectrum ($S/N < 5$), we view neither AstroDash's nor SNID's classification as particularly reliable.   These results enhance the importance of using  multiple methods to perform spectral classification.

\subsection{Spectral fitting with { KN} models} \label{sec:Kasen}
{ KNe} are expected to produce quasi-blackbody radiation. They are expected to have a rapidly changing lightcurve, a luminosity consistent with nuclear rapid neutron capture (r-process) heating, and a long-lived infrared emission. Analysis of the spectrum of AT2017gfo (the 
{ KN} associated with GW170817) showed emission from both light r-process and heavy r-process components which led to a spectrum that appears as a superposition of two blackbodies at different temperatures. At early times the spectra are mostly featureless, while at later times there are distinct features in the infrared. 

For our analysis, we used the set of synthetic kilonova spectra by \citet{kasen_2017} ({ see} \S~\ref{sec:software}).  This set of \citet{kasen_2017} models covers a regularly sampled grid in parameter space of ejecta mass ($M = 0.001 - 0.1 M_{\sun}$), ejecta velocity ($v_{\rm kin} = 0.03 - 0.40c$), and ejecta lanthanide mass fraction ($X_{\rm lan} = 10^{-9} - 10^{-1}$).  At each of these grid points in ($M$,$v_{\rm kin}$,$X_{\rm lan})$-space is a time series of synthetic spectra spaced in units of 0.1~day from $\approx$2 days pre-merger out to $\approx$25 days post-merger.  Each of these synthetic spectra covers a rest-frame wavelength range from the ultraviolet ($\approx$ 150\AA) through the infrared ($\approx$ 10$\mu$m).   

We took the processed and calibrated observed spectrum for each of our 
{ KN} candidates and performed a least-squares fit to the \citet{kasen_2017} grid of synthetic spectra for the appropriate time post-merger when the candidate's spectrum was observed.    
In this fit, the redshifts of the synthetic spectra were also allowed to float within a 1$\sigma$ range centered on the estimated redshift of the LVC source 
($z=0.059\pm0.011$), 
yielding a best-fit spectrum at a best-fit redshift.

In Figure \ref{fig:thumbnails-all4} --
\ref{fig:thumbnails-all11} we show the results of these fits for our sample of observed 
{ KN} candidate spectra.
With the possible exception of AT2019ntr, none of these candidates have an 
observed spectrum that is a particularly good fit to the \citet{kasen_2017} models -- mostly due to the appearance of one or more strong emission features in the observed spectrum --  which is consistent with our conclusion that none of these objects is a 
{ KN}, but rather each is { an SN} from one of several types.  What of AT2019ntr? For this object the best-fit redshift ($z_b=0.049$) is on the low end, but still within the $1\sigma$ errors from the redshift based on the original LVC O3 distance estimate ($z=0.059\pm0.011$).  Furthermore, this is one of the cases where the AstroDash and SNID fits are both poor (low {\texttt rlap}) and inconsistent with each other (see Table~\ref{tab:candidates3}). So, is AT2019ntr the optical counterpart to GW190814?  Unfortunately, we cannot provide a definite conclusion based on the SOAR data alone.  As it turns out, though, it is unlikely that AT2019ntr is the 
{ KN} we were seeking:  first, its sky coordinates {lie outside the final LVC 90\% confidence contour} for GW190814 (see Fig.~\ref{fig:prob_map}); secondly and more importantly, in their analysis of the DECam data for these candidates, \citet{Morgan_paper} applied a light-curve-based machine (ML) classifier -- a combination of \citet{2011ApJ...738..162S}'s {\texttt{PSNID}} fitting code and a random forest classifier -- to the photometric time series data for AT2019ntr, and this yielded a 96\% probability that AT2019ntr is { an SN}.

Finally, it might be asked whether it would not be more efficient to add the Kasen templates into AstroDash/SNID so one could directly compare the likelihood that an object is a classical SN vs.\ a KN.  
One of the first things AstroDash/SNID does is to fit the continuum of the spectrum and remove it.  KN spectra -- especially early on in their light curves -- 
are continuum dominated, with few prominent emission/absorption features.   Thus, there would be 
little left to fit in the case of the KNe models.  Maybe a version of AstroDash/SNID that did {\em not\/} subtract off the continuum during the fit would work, but that would be a future project.   

\subsection{Spectra of Host Galaxies} \label{sec:host_galaxies}

Finally, there were three candidates which were too faint for us to target effectively with SOAR (AT2019nte, AT2019omw, AT2019omx).  We instead targeted the host galaxy, with the idea that, if the host galaxy's redshift was significantly discordant with that of the {distance estimated from the GW signal, that would rule out that candidate as a possible counterpart to GW190814.}  We found that only one (AT2019omx) had a truly discordant redshift ($z=0.275$); see Figure~\ref{fig:thumbnails-all1}.  The host galaxies of the other two candidates, AT2019nte ($z=0.070$; Fig.~\ref{fig:thumbnails-all2}) and AT2019omw ($z=0.047$; Fig.~\ref{fig:thumbnails-all3}) have redshifts that are consistent with the {redshift corresponding to the GW distance} at about the 1$\sigma$ level.  As it turns out, in the end both AT2019nte and AT2019omw failed the DESGW Search \& Discovery {\em offline} imaging pipeline criteria for a good candidate:  AT2019nte because it did not meet a sufficiently high detection threshold in the DECam imaging, and AT2019omw because it did not survive the offline visual inspection of candidates \citep{Morgan_paper}.  Thus, we consider all three of these candidates as being ruled out.

\subsection{Lessons Learned from DESGW Spectroscopy in O3} \label{sec:lessons}

One of final results we would like to discuss are those of ``lessons learned'' during the concerted effort by the { DESGW} imaging and spectroscopic follow-up teams during the follow-up of GW190814 candidates, particularly as the spectroscopic follow-up of this LVC event may be viewed as a template for future spectrosopic follow-ups in LVC O4 and beyond, since, 
as the LVC becomes increasingly more sensitive, the optical counterparts of future LVC events will likely be relatively distant and faint, unlike the very nearby and bright BNS 
{ KN} GW170817.

First, we found that our SOAR spectroscopic follow-up effort benefited from being a loose confederation of semi-independent teams that could operate the telescope remotely:  a team based at Fermilab, a team based at University of California - Santa Cruz, a team based in Chile, and a team based in Brazil.  Each of these teams signed up to be ``on-call'' for 2-week blocks throughout LVC O3.  The team ``on-call'' when an LVC O3 alert went out would have the responsibility for preparing and carrying out any SOAR spectroscopic follow-up during their watch.  That said, the ``on-call'' team could request help from the other teams, and
the other teams were welcome to follow along during the night of a follow-up observation.  In the case of GW190814, the Fermilab team was the on-call team for most of the time of the spectroscopic follow-up, but other teams also provided help during Fermilab's time block (in particular, the Chilean team took over a couple nights when the Fermilab team was unable to observe).  This relatively loose structure of our spectroscopic follow-up effort seemed to work well, especially over the full course of LVC O3.

Second, especially as SOAR is primarily run as a remote observing facility, it is vital to have good communications with the SOAR scientific and technical staff. We were able to easily communicate with the SOAR staff and on several occasions SOAR staff provided invaluable help to us in obtaining spectra of dimmer objects that required a longer process for target acquisition.  Further, long after the optical signature of any expected 
{ KN} should have faded, the SOAR staff obtained the spectra of the host galaxies of two remaining candidates (AT2019nte and AT2019omw) during engineering time, in order to check if these candidates had redshifts that fell within the distance estimates measured by LVC for the GW event.

Third, it became clear early on that it is very difficult to obtain sufficiently high $S/N$ spectra with SOAR for candidate 
{ KNe} fainter than about $i\approx21$ in the allotted time for a SOAR ToO interrupt. 
For spectroscopic follow-up in LVC O4, candidates fainter than $i\approx21$ should either be pursued by 6-to-10-meter-class telescopes, or have their host galaxies targeted as a means to qualify them or to rule them out.

Finally, we stress the importance of being able to reduce and analyze the data at the telescope for quick classification of the candidate as a 
{ KN} or not.  If there are obvious features in the spectrum indicating that a given candidate is not a 
{ KN} (e.g., sharp emission or absorption lines or features typical of { an SN} spectrum), one can quickly move on to the next target in the candidate list; if, however, the spectrum indicates that the candidate is indeed the 
{ KN}, the rest of the astronomical community can be quickly alerted.  At the telescope during the observations for this paper, we typically made use of our 
SOAR Quick Reduce Pipeline or IRAF routines to process and calibrate the spectra on the fly, and classified the spectra by eye or by running them through the AstroDash and/or the SNID { SN} typing software that same night.
A later, more refined reduction and analysis were performed later offline, as described in \S~\ref{sec:spectral} and \S~\ref{sec:reliability}.  
We note that, however, whereas some of the classifications changed between the real-time and off-line analysis, none of the resulting spectra { -- with the possible exception of the very low-$S/N$ AT2019ntr spectrum -- } were ever { seriously} considered { to be} that of a KN:  {\em i.e.\/}, the quick reductions are sufficient for the purpose. One weakness during our O3 observations of GW190814 candidates was the lack of an analog of our Quick Reduce pipeline to fit a candidate's spectrum to a grid of 
{ KN} model spectra on the fly at the telescope.  Since then, we have developed an initial version of own { publicly available} DESGW
{ KN} spectrum fitter (DLT\_DESGW\_KNfit; see  \S~\ref{sec:software}), which can be run at the telescope with the output of our SOAR Quick Reduce pipeline and should be useful for spectroscopic follow-up in LVC O4.

\section{Conclusions}
\label{sec:conclusions}
{ In the era of multi-messenger astronomy, we have demonstrated that we can perform a deep, one-of-its-kind spectroscopic follow-up campaign for possible NSBH events.} 
We have reported on the SOAR/Goodman spectroscopy of 11 
{ KN} candidates associated with the LIGO/VIRGO event GW190814. For 8 of these we have reported the redshift and spectroscopic typing of the transient itself, and for the other 3 we have reported the redshift of the host galaxy. We concluded that none of these candidates were the optical counterpart associated with the compact object binary merger. This SOAR/Goodman spectroscopy was done through SOAR ToO
observations on a series of nights following the LVC discovery of gravitational waves from GW190814. These targeted observations were performed after 
{ KN} candidate identification and culling by the DESGW collaboration following observations using DECam on the Blanco telescope, and they have allowed us to place interesting constraints on the properties of the binary \citep{Morgan_paper} and to use this event as a dark standard siren (that is, as a constraint on $H_0$ using { GWs}) \citep{darksiren2}.

We have also described the DESGW spectroscopic pipeline, part of the DESGW 
{ KN} search process and candidate assessment, and our process and timeline for creating a spectroscopic follow-up candidate list.  In addition, we have presented our QuickReduce software (for quick look spectroscopic reduction) and the UCSC Reduction Pipeline software (for offline spectroscopic reduction).  Furthermore, we have shown our use of AstroDash, SNID, and a least-square KN model fitting software for the process of candidate spectrum classification. Finally, we have demonstrated the effectiveness of our program and these tools within DESGW and are prepared for more extensive searches for 
{ KNe} in LVC O4.

\section{Software}
\label{sec:software}
We present here links to the software packages mentioned in the text:
\vspace{-0.2cm}
\begin{enumerate}
\item Quick Reduce Pipeline, used for reduction and analysis of spectra immediately after observing. \url{https://github.com/DouglasLeeTucker/SOAR_Goodman_QuickReduce/blob/master/notebooks/SOAR_Goodman_QR_Notebook.ipynb}
\vspace{-0.2cm}
\item UCSC spectral pipeline, used for data reduction and analysis: \url{https://github.com/msiebert1/UCSC_spectral_pipeline}
\vspace{-0.2cm}
\item AstroDash supernova typing software: \url{https://github.com/daniel-muthukrishna/astrodash}
\vspace{-0.2cm}
\item Image Reduction and Analysis Facility (IRAF). IRAF { had been} distributed by the National Optical Astronomy Observatory, which { was} operated by the Association of Universities for Research in Astronomy (AURA) under a cooperative agreement with the National Science Foundation.  { The software is currently maintained and distributed by the IRAF Community: \url{https://iraf-community.github.io/}}
\vspace{-0.2cm}
\item SNID supernova typing software: \url{https://people.lam.fr/blondin.stephane/software/snid/}
\vspace{-0.2cm}
\item Kasen 
{ KN} models: \url{https://github.com/dnkasen/Kasen_Kilonova_Models_2017}
\vspace{-0.2cm}
\item DESGW 
{ KN} spectrum fitting software:
\url{https://github.com/cdebom/DLT_DESGW_KNfit}
\vspace{-0.2cm}
\item SNANA SuperNova ANAlysis software
\url{https://snana.uchicago.edu/}
\vspace{-0.2cm}
\item matplotlib \citep{2007CSE.....9...90H}, 
\vspace{-0.2cm}
\item numpy \citep{numpy:2011}, 
\vspace{-0.2cm}
\item scipy \citep{scipy:2001}, 
\vspace{-0.2cm}
\item astropy \citep{Astropy:2013}, 
\vspace{-0.2cm}
\item TOPCAT \citep{2005ASPC..347...29T}.
\vspace{-0.2cm}
\end{enumerate}

\section*{Acknowledgments}
Funding for the DES Projects has been provided by the DOE and NSF(USA), MEC/MICINN/MINECO (Spain), STFC (UK), HEFCE(UK). NCSA (UIUC), KICP (U. Chicago), CCAPP (Ohio State), 
MIFPA (Texas A\&M), CNPQ, FAPERJ, FINEP (Brazil), DFG (Germany) and the Collaborating Institutions in the Dark Energy Survey.

The Collaborating Institutions are Argonne Lab, UC Santa Cruz, University of Cambridge, CIEMAT-Madrid, University of Chicago, University College London, 
DES-Brazil Consortium, University of Edinburgh, ETH Z{\"u}rich, Fermilab, University of Illinois, ICE (IEEC-CSIC), IFAE Barcelona, Lawrence Berkeley Lab, 
LMU M{\"u}nchen and the associated Excellence Cluster Universe, University of Michigan, NOAO, University of Nottingham, Ohio State University, University of 
Pennsylvania, University of Portsmouth, SLAC National Lab, Stanford University, University of Sussex, Texas A\&M University, and the OzDES Membership Consortium.

Based in part on observations at Cerro Tololo Inter-American Observatory, National Optical Astronomy Observatory, which is operated by the Association of 
Universities for Research in Astronomy (AURA) under a cooperative agreement with the National Science Foundation.

The DES Data Management System is supported by the NSF under Grant Numbers AST-1138766 and AST-1536171. The DES participants from Spanish institutions are partially 
supported by MINECO under grants AYA2015-71825, ESP2015-88861, FPA2015-68048, and Centro de Excelencia SEV-2016-0588, SEV-2016-0597 and MDM-2015-0509. Research leading 
to these results has received funding from the ERC under the EU's 7$^{\rm th}$ Framework Programme including grants ERC 240672, 291329 and 306478.

We acknowledge support from the Australian Research Council Centre of Excellence for Gravitational Wave Discovery (OzGrav) project CE170100004.

The UCSC team is supported in part by NASA grant NNG17PX03C, NSF grant AST-1815935, the Gordon \& Betty Moore Foundation, the Heising-Simons Foundation, and by a fellowship from the David and Lucile Packard Foundation to R.J.F.

IA is a CIFAR Azrieli Global Scholar in the Gravity and the Extreme Universe Program and acknowledges support from that program, from the European Research Council (ERC) under the European Union’s Horizon 2020 research and innovation program (grant agreement number 852097), from the Israel Science Foundation (grant number 2752/19), from the United States - Israel Binational Science Foundation (BSF), and from the Israeli Council for Higher Education Alon Fellowship.

DAH is supported by NSF grant AST-1911151.

R. Morgan thanks the LSSTC Data Science Fellowship Program, which is funded by LSSTC, NSF Cybertraining Grant \#1829740, the Brinson Foundation, and the Moore Foundation; his participation in the program has benefited this work.

FOE acknowledges support from FONDECYT grant 1201223.

L. Santana-Silva acknowledges the financial support from FAPESP through the grant $\#2020/03301-5$.

Based on observations obtained at the Southern Astrophysical Research (SOAR) telescope, which is a joint project of the Minist\'erio da Ci\^encia, Tecnologia e Inova\c c\~ao (MCTI) da Rep\'ublica Federativa do Brasil, the U.S. National Optical Astronomy Observatory (NOAO), the University of North Carolina at Chapel Hill (UNC), and Michigan State University (MSU).

This research uses services or data provided by the NOAO Science Archive. NOAO is operated by the Association of Universities for Research in Astronomy (AURA), Inc. under a cooperative agreement with the National Science Foundation.

This manuscript has been authored by Fermi Research Alliance, LLC under Contract No. DE-AC02-07CH11359 with the U.S. Department of Energy, Office of Science, Office of High Energy Physics.  The U.S. Government retains and the publisher,
by accepting the article for publication, acknowledges that the U.S. Government retains a non-exclusive, paid-up, irrevocable, world-wide license to publish or reproduce the published form of this manuscript, or allow others to do so,
for U.S. Government purposes.

\bibliographystyle{yahapj}
\bibliography{bibliography}

\end{document}